\documentclass[fleqn,usenatbib]{mnras}

\usepackage{graphicx}	% Including figure files
\usepackage{amsmath}	% Advanced maths commands
\usepackage{amssymb}	% Extra maths symbols
\usepackage{adjustbox}
\usepackage{threeparttable}
\usepackage{tabto}
\usepackage{float}
\usepackage{color}
\usepackage{graphics}
\usepackage{adjustbox}
\usepackage[usenames,dvipsnames]{pstricks}
\usepackage{epsfig}
\usepackage{pst-grad} % For gradients
\usepackage{pst-plot} % For axes
\usepackage{times,pstricks,pst-node,pst-coil}
\usepackage{multicol} 
\usepackage{multirow} 
\usepackage{indentfirst} 
\usepackage{natbib} 
\usepackage{threeparttable}
\usepackage{longtable}
\usepackage{pdflscape}
\usepackage{pdfpages}

\newcommand{\Msun}{M$_{\odot}$}

\newcommand{\bse}{{\sc bse}}

\definecolor{smalt(darkpowderblue)}{rgb}{0.0, 0.2, 0.6}
\definecolor{forestgreen(traditional)}{rgb}{0.0, 0.5, 0.0}
\defcitealias{WW02}{WW02}
\defcitealias{LWW94}{LWW94}
\newcommand{\ww}{\citetalias{WW02}}

\newcommand{\lww}{\citetalias{LWW94}}

\usepackage{etoolbox}
\makeatletter
\patchcmd\@combinedblfloats{\box\@outputbox}{\unvbox\@outputbox}{}{%
   \errmessage{\noexpand\@combinedblfloats could not be patched}%
}%
 \makeatother

\title[Evidence for reduced MB in polars from BP models]
{Evidence for reduced magnetic braking in polars from binary population models}

\author[D. Belloni et al.]{
Diogo Belloni$^{1,2}$\thanks{diogo.belloni@inpe.br (DB)},
Matthias R. Schreiber$^{2,3}$\thanks{matthias.schreiber@uv.cl (MRS)},
Anna F. Pala$^{4}$,
Boris T. G\"ansicke$^{5}$,
\newauthor 
M\'onica Zorotovic$^{2}$ and
Claudia V. Rodrigues$^{1}$\\
$^{1}$National Institute for Space Research, Av. dos Astronautas, 1758, 12227-010, S\~ao Jos\'e dos Campos, SP, Brazil\\
$^{2}$ Instituto de F{\'i}sica y Astronom{\'i}a, Universidad de Valpara{\'i}so, Av. Gran Breta{\~n}a 1111, 2360102, Valpara{\'i}so, Chile \\
$^{3}$ Millenium Nucleus for Planet Formation, Universidad de Valpara{\'i}so, Av. Gran Breta{\~n}a 1111, 2360102, Valpara{\'i}so, Chile\\
$^{4}$European Southern Observatory, Karl Schwarzschild Stra{\ss}e 2, Garching, 85748, Germany\\
$^{5}$ Department of Physics, University of Warwick, Coventry CV4 7AL, UK
}

\date{Accepted: 2019 December 02; Revised: 2019 November 27; Received: 2019 October 09}

\pubyear{2019}

\begin{document}
\label{firstpage}
\pagerange{\pageref{firstpage}--\pageref{lastpage}}
\maketitle

\begin{abstract}
We present the first population 
synthesis of 
synchronous magnetic
cataclysmic variables, 
called polars, taking 
into 
account the effect of the white dwarf (WD) magnetic 
field on angular momentum loss.
We implemented the reduced 
magnetic braking (MB) model
proposed by Li, Wu \& Wickramasinghe into 
the Binary Stellar Evolution ({\sc bse}) code
recently calibrated 
for cataclysmic variable (CV) evolution.
We then compared separately our predictions for polars and non-magnetic CVs 
with a large and homogeneous sample of observed CVs from
the \emph{Sloan Digital Sky Survey}. 
We found that the predicted orbital period distributions and space densities  
agree with the observations if period bouncers are excluded. 
For polars, we also find agreement between predicted and observed mass transfer rates, 
while the mass transfer rates of non-magnetic CVs 
with periods ${\gtrsim3}$~h drastically disagree with those 
derived from observations. 
Our results provide strong evidence that the
reduced MB model for the evolution of highly 
magnetized accreting WDs can 
explain the observed properties of polars.
The remaining main issues in our understanding 
of CV evolution are 
the origin of the large number of highly magnetic WDs, 
the large scatter of the observed mass transfer rates 
for non-magnetic systems with periods ${\gtrsim3}$~h, 
and the absence of period bouncers in observed samples.
\end{abstract}

% Select between one and six entries 
%from the list of approved keywords.
% Don't make up new ones.

%%%%%%%%%%%%%%%%
%%%%%%%%%%%%%%%%
%%%%%%%%%%%%%%%%
% NEW SECTION
%%%%%%%%%%%%%%%%
%%%%%%%%%%%%%%%%
%%%%%%%%%%%%%%%%
\section{Introduction}

Cataclysmic variables 
(CVs) are interacting binaries 
composed of a white dwarf (WD) that accretes 
matter from a low-mass donor star
\citep[e.g.][for comprehensive reviews]
{Warner_1995_OK, Hellier_2001}. 
Because of their rich variety of variabilities 
(basically over all wavelengths and on 
a wide range of time-scales) caused by 
different physical processes, studying CVs bears
relevance for several fields of research  
including binary formation and evolution, 
accretion processes, supernova Ia progenitors,
and interactions between dense plasma 
and very strong magnetic fields.

CVs can be classified according to the 
WD magnetic field strength. 
Non-magnetic CVs are systems in which  
the magnetic field is negligible 
leading to accretion via a disc,
which is formed due 
to angular momentum conservation during 
mass transfer.
Magnetic CVs are systems where 
the accretion is partially or totally
along the WD magnetic field lines.
Given that
roughly one-third of known CVs in 
the solar neighborhood are magnetic
\citep{Pala_2019b},
it is alarming that 
thorough population synthesis
of CVs, in which the WD magnetic field
is properly taken into account,
has never been carried out.

Magnetic CVs are divided into two 
main classes,
intermediate polars (IPs)
and polars
\citep{Cropper_1990,
Patterson_1994,
Wickramasinghe_2000,
Ferrario_2015}.
The main difference between these 
two types is that in polars, not only the
donor spin, but also the WD spin 
are synchronised with the orbit, presumably 
because the WD magnetic field exerts 
a synchronising torque on the donor. 
In addition, polar WD magnetic fields 
are of the order of {${10^7-10^8}$\,G},
which prevents the formation of an
accretion disc.
In IPs, the WD spin is not synchronised 
with the orbit because 
the WD magnetic field ({$\sim10^6-10^7$\,G}) 
is not strong enough,
which implies that
most IPs possess truncated
accretion discs.

With respect to CV evolution 
\citep[e.g.][]{Knigge_2011_OK}, non-magnetic
systems are expected to evolve towards 
shorter periods due to angular momentum 
loss (AML) caused mainly by 
magnetic braking (MB) and 
gravitational 
radiation (GR).
In the standard CV evolution model,
MB dominates the AML above the orbital
period gap (orbital periods longer 
than $\approx3$~h, e.g.
\citealt{Rappaport_1982}), 
while GR is expected to drive the evolution 
in CVs with shorter periods
\citep[e.g.][]{Paczynski_1967}.
The orbital period gap in the standard
model is explained via the disrupted MB
scenario, which has gained considerable 
support in the past few years
\citep{schreiberetal10-1,Zorotovic_2016}.
MB causes mass transfer rates
high enough to
drive the main-sequence (MS) donor out of 
thermal equilibrium, leading to a bloated 
MS star (about 30 per cent larger
in radius). 
When the donor star becomes fully 
convective at an orbital period of 
$\approx3$~h, MB is 
expected to cease
(or become much less efficient), 
which leads to a significant drop in the
mass transfer rate and a slowdown of
the evolution, as the system is now
driven by GR only.
Such a drop in the mass transfer rate 
allows the MS donor to re-establish
thermal equilibrium (i.e. decrease in size), 
and eventually the system becomes a detached binary  
since the secondary is no longer filling its 
Roche lobe.
Even though the system is now detached, 
it 
keeps 
loosing angular momentum due
to GR and continues to evolve towards shorter 
orbital periods.
When the orbital period is $\approx2$~h, 
the MS donor fills its Roche lobe again,
and mass transfer restarts, i.e. the system 
is again a CV evolving now with relatively
low mass transfer rates towards 
shorter orbital periods.
When the period is 
$\approx80$~min, the mass loss rate from
the secondary drives it increasingly out of 
thermal equilibrium and it becomes a 
hydrogen-rich degenerate object.
From this point, the donor 
expands in response to the mass loss, 
leading to an increase in the orbital period.
At this phase,
the CV is called a period bouncer.
For decades, population models 
of non-magnetic CVs have been 
developed, which substantially improved  
our understanding of \emph{non-magnetic} CV 
evolution
\citep[e.g.][]
{Rappaport_1982,
Kool_1992,
Kolb_1993,
Politano_1996,
Howell_2001,
Knigge_2011_OK,
Goliasch_2015,
Kalomeni_2016,
Schreiber_2016,
Belloni_2018b}.

The evolution of polars, however, is most 
likely different to those of non-magnetic CVs. 
In these systems the WD magnetic field is expected 
to affect MB. 
There are two main evolutionary 
models that address changes in MB 
in polars, namely 
the {\it enhanced MB} 
\citep*[e.g.][]{Hameury_1989} and 
the {\it reduced MB} 
\citep*[e.g.][]{LWW94,WW02} models. 
Both models
assume a coupling of the 
donor magnetic field lines and 
the WD magnetic field lines.
In the enhanced MB scenario, 
the wind from 
the MS donor is
trapped by the WD 
magnetic field lines, 
which would increase 
the magnetosphere radius
resulting in enhanced AML via MB.
This would cause an enhanced
mass transfer, leading to correspondingly 
shorter evolutionary time-scales. In many cases,
mass transfer could even be thermally
unstable, which would result in a very short 
life-time for such systems.
In contrast, 
in the reduced MB scenario, 
winds from the MS donor 
do not carry away as much angular 
momentum as in the
non-magnetic case
resulting in reduced AML via MB.
This is because part of the wind
remains trapped to the system due
to the strong WD magnetic field.
Whether the observed population 
of polars, in particular their 
orbital period distribution, is 
consistent with 
either the enhanced or the reduced MB models,
has never been tested.

To progress with this situation, we present 
here realistic population models 
of polars.
We incorporate the reduced MB scenario in an existing binary population code,   
and investigate how the WD magnetic field affects CV evolution. 
We focus on the reduced
MB model only
because the mass transfer rates in 
polars derived from observations are smaller than in 
non-magnetic CVs
\citep{Araujo_2005,TG09}. These measurements are
consistent with reduced AML but clearly contradict enhanced AML. 
Having incorporated reduced magnetic braking 
in our CV evolution model, we compare the model predictions with 
observed orbital periods and
mass transfer rates as well as with estimates of the space density.

%%%%%%%%%%%%%%%%
%%%%%%%%%%%%%%%%
%%%%%%%%%%%%%%%%
% NEW SECTION
%%%%%%%%%%%%%%%%
%%%%%%%%%%%%%%%%
%%%%%%%%%%%%%%%%
\section{Impact of the WD Magnetic Field on Magnetic Braking}
\label{mag}

In order to account for the influence 
of the WD magnetic field $(B_{\rm WD})$ 
in our simulations, we adopted
here the formulation by 
\citet[][hereafter WW02]{WW02},
which is in turn based on the 
\textit{reduced MB} model 
proposed by 
\citet*[][hereafter LWW94]{LWW94}.

In a rotating M-dwarf on the 
main sequence with convective 
envelope, a dynamo process is 
able to generate a surface magnetic field 
\citep[e.g.][]
{Schatzman_1962,Weber_1967,Mestel_1968}.
If this surface magnetic field is 
sufficiently strong, it can exert 
magnetic torques on winds 
such that co-rotation is established, 
and the outflowing material follows the magnetic 
field lines.
This way, winds can carry off 
a substantial amount of
angular momentum per unit mass 
\citep[e.g.][]{Mestel_1987}.
This is the basic mechanism driving 
MB. The resulting AML proportionally 
depends on the stellar rotation 
(i.e. the greater the star 
rotation, the greater the 
wind-driven AML).

More detailed models show, however, that 
only part of the flow escapes
from the star's surface
and that one can separate two  
regions, a 
{\it dead zone}
 and a 
{\it wind zone} 
\citep[e.g.][see fig.~1]{Mestel_1968}. 
The dead zone corresponds to the 
region where gas is prevented from 
escaping the star due to the 
pressure of closed magnetic 
field loops, i.e. the magnetic 
pressure is greater than the thermal
pressure. 
The wind zone 
is the region where the field lines 
are open and gas 
escapes and carries off angular momentum.

In the case of synchronously rotating 
CVs (i.e. polars),
it is quite likely that $B_{\rm WD}$
affects the interplay 
between the donor wind and dead zones, 
as it is strong enough to synchronize 
both stars of the binary system.
\lww~developed a model for MB in 
which the donor wind zone is reduced 
due to the formation of additional closed 
field lines, which connect the escaping gas 
directly to the WD. 
This results in the 
gas following the WD magnetic field 
lines, instead of the open 
M-dwarf magnetic field lines.
This causes a reduction
of the wind zone, by creating a 
second dead zone 
\citepalias[][their~fig.~1]{LWW94}.

In the model proposed by \lww, 
the basic assumptions are: 
(i) both WD and donor have 
centred dipole magnetic fields;
(ii) the donor magnetic moment 
is oriented perpendicular to the 
orbital plane;
(iii) the WD magnetic moment 
$(\mu_{\rm WD})$ is anti-aligned 
with that of the donor; 
(iv) the total gas flux carried 
by open field lines is conserved 
from the donor to
the Alfv\'en surface.
Because of these strong assumptions,
the model is clearly simplistic. 
For instance, $\mu_{\rm WD}$ 
could have in principle any orientation
and \citet{Li_1998} showed that
detailed evolutionary modelling of 
polars needs to take into account 
inclination effects.
In addition, the magnetic field 
may have complex
topologies so 
that adopting dipolar fields might not
be adequate.
However,
it is the only currently available model that can
be incorporated in binary population
synthesis codes and we test here if it is sufficiently realistic 
to explain the long-term evolution of polars.

In order to find a description of the limiting lines
separating the dead zones and the 
wind zone, \lww \, assume an 
isothermal gas and solve the 
magnetohydrodynamical
equations in the dead zones. 
Based on these solutions, they could estimate 
the reduction of MB 
for a given $B_{\rm WD}$.
In \lww, the results are parametrized 
with the
{\it fraction of open field lines $(\Phi)$}.
In the extreme case of
no wind zone, i.e. $B_{\rm WD}$ is 
strong enough to prevent any MB, 
$\Phi=0.0$. 
On the other hand, in the case of 
negligible $B_{\rm WD}$, the second
dead zone does not exist and 
the MB is not 
reduced, $\Phi = 0.258$.
This prescription for the reduction of MB  
parametrized 
with $\Phi$ (which depends on the 
binary parameters and $B_{\rm WD}$) allows to incorporate 
the impact of strong WD magnetic fields 
in binary evolution codes.

It has been shown by \ww \, that the
AML due to MB in polars 
$(\dot{J}_{\rm MB,pol})$
takes rather a simple form 
when the donor wind is centrifugally 
driven: 

\begin{equation}
\dot{J}_{\rm MB,pol} \ = \ 
\dot{J}_{\rm MB,non-mag} \, \left( \frac{\Phi}{0.258} \right)^{5/3} \, ,
\label{EQN1}
\end{equation}
\

\noindent
where 
$\dot{J}_{\rm MB,non-mag}$ is the
AML due to MB in non-magnetic CVs.
In this work we used the prescription from
\citet{Rappaport_1983} with $\gamma=3$, i.e.

\begin{equation*}
\dot{J}_{\rm MB,non-mag} \ = 
\ -5.83 \times 10^{-16}  \, \left[ M_2 \left( R_2 \, \Omega_2 \right)^3 \right] \ \ 
{\rm M}_\odot \, {\rm R}_\odot^2 \, {\rm yr}^{-2} \, , \label{EQN2}
\end{equation*}
\

\noindent
where 
$M_2$ and $R_2$ are the donor mass 
and radius, respectively, both
in solar units, and $\Omega_2$ is 
the donor spin in units of year.

In the standard scenario for CV evolution, the donor MS star becomes 
significantly bloated as a response to
mass transfer in non-magnetic CVs above the orbital period gap.  
In the model described here, the strongly magnetic WD 
reduces MB and thus the mass transfer rate in polars. 
We therefore also expect the secondary to be less bloated in polars.
This implies that for polars
the critical mass at which MB
is disrupted is larger 
than the critical value of 
$M_2^{\rm crit,non-mag}=0.2$~\Msun~
found for non-magnetic CVs. 
In the limit of full MB suppression,
one would expect
that the critical mass is precisely
the same as for isolated stars
or MS stars in detached binaries, 
i.e. $M_2^{\rm crit,sgl}=0.35$~\Msun~
\citep[e.g.][]{Reiners_09,schreiberetal10-1}.

If we assume the same dependence
on $\Phi$ that has been derived for the angular momentum loss (Eq. \ref{EQN1})
for the critical donor mass 
in polars at which
MB is disrupted, $M_2^{\rm crit,pol}$,  
and for the radius increase due to mass loss, we find relatively simple 
relations for the donor radius and 
the mass limit of angular momentum loss through MB. 
The critical mass is then given by

\begin{equation}
M_2^{\rm crit,pol} = M_2^{\rm crit,sgl} \ - \ \left( M_2^{\rm crit,sgl} \ - \ M_2^{\rm crit,non-mag} \right) \, \left( \frac{\Phi}{0.258} \right)^{5/3} \, .
\label{EQN3}
\end{equation}
\

\noindent
and the expansion factor of the 
MS donor as a function of $\Phi$
in the case MB is reduced is given by
\begin{equation}
f_{\rm bloat,pol} =  f_{\rm bloat,non-mag} \, \left( \frac{\Phi}{0.258} \right)^{5/3} \, ,
\label{EQN4}
\end{equation}
\

\noindent
where $f_{\rm bloat,non-mag} \approx 0.357$ \citep[e.g.][]{Davis_2008}.
This means that in polars the donor radius  
is approximated as 
\begin{equation}
R_2^{\rm pol} \ = \ R_2^{\rm sgl} \, \left( \,  1 \ + \ f_{\rm bloat,pol} \, \right) \, ,
\label{EQN5}
\end{equation}
\

\noindent
where $R_2^{\rm sgl}$ is the 
radius of MS single stars or MS stars in detached binaries.
\

It is important to note that 
all quantities in Eqs. 
\ref{EQN1}, \ref{EQN3},
\ref{EQN4} and \ref{EQN5} 
are reduced to the standard CV 
evolution model in case of {${\Phi\geq0.258}$} 
(non-magnetic WD). 
Instead, in the case of extremely high $B_{\rm WD}$, 
{${\Phi=0.0}$} (highly magnetic WD) our formulation leads to a 
CV evolution driven only by GR. In this case,  
the bloating of the donor will negligible and
therefore such systems will not evolve through a detached phase, 
i.e. for them there will 
be no orbital period gap.

\section{The assumed white dwarf magnetic
field topology and strengths}
\label{obsBwd}

For our modelling of reduced MB in polars, 
we emphasize that in our approach, as in the case of \ww, 
we assume that the WD is synchronized 
with the orbit since the onset of 
mass transfer, i.e. that the system
is a polar throughout its evolution with a constant magnetic field strength.

\begin{figure}
   \begin{center}
    \includegraphics[width=0.975\linewidth]
    {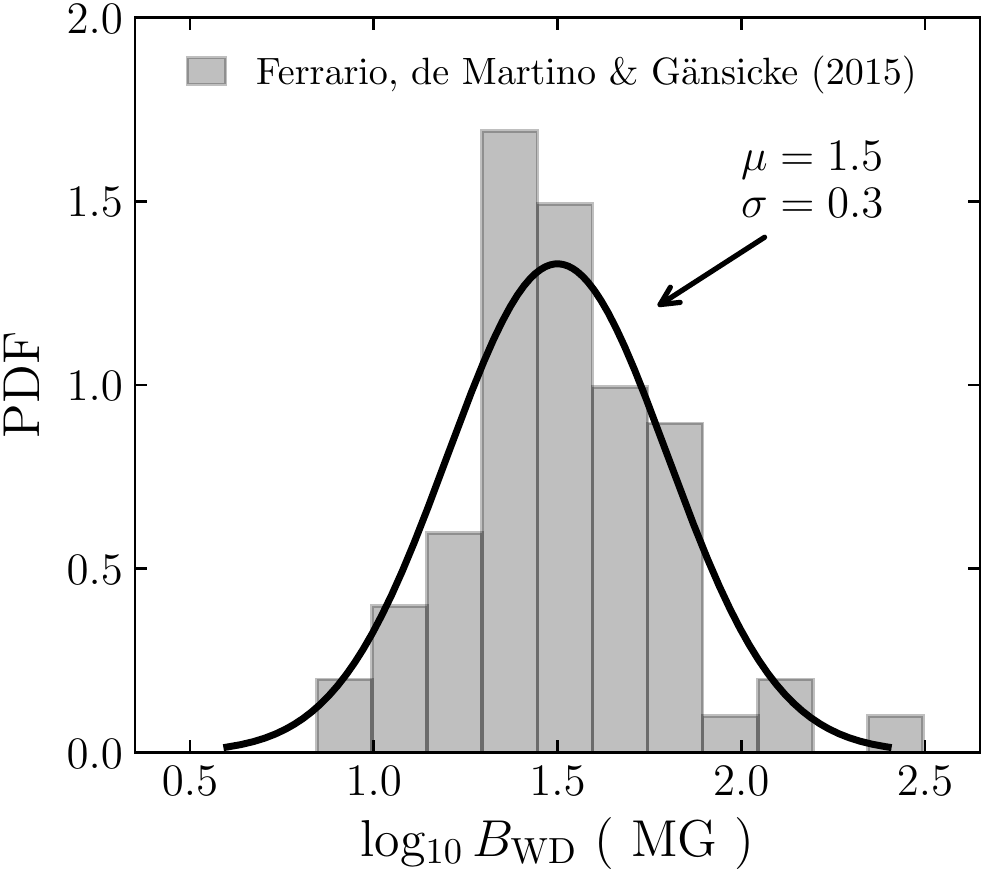}
    \end{center}
  \caption{Observed distribution 
  of WD magnetic field in polars,
  from \citet*[][table 2]{Ferrario_2015}.
  The line corresponds to the 
  best-fitting Gaussian
  $(\mu=1.5;~\sigma=0.3)$,
  which is used to obtain the
  WD magnetic fields in the 
  simulations.
  }
  \label{Fig01}
\end{figure}

This approach is justified as we are primarily interested 
in understanding how an existing $B_{\rm WD}$ 
affects CV evolution and 
the predicted period distribution.  
We therefore used the 
observed sample of polars which contains 
67 polars with measured $B_{\rm WD}$ as listed in 
\citet*[][table 2]{Ferrario_2015}.
The corresponding $B_{\rm WD}$ distribution is shown in
Fig.~\ref{Fig01}. 
From this observed $B_{\rm WD}$ distribution, 
we obtained the best-fitting Gaussian to
the $\log_{10}B_{\rm WD}$ distribution,
which has mean and standard deviations
given by $1.5$ and $0.3$, respectively.
As the true distribution might not be a Gaussian, 
we performed several tests using different 
probability density functions derived from the observed distributions, 
such as the kernel density estimation method,
and found that the conclusions of this paper 
are independent of the detailed modelling of the data.

The WD magnetic moment of a given polar in our 
simulations in given by

\begin{equation}
\mu_{\rm WD} \ = \ \frac{1}{2} \, 
\left( \, \frac{B_{\rm WD}}{{\rm G}} \right) \, \left( \frac{R_{\rm WD}}{{\rm cm}} \right) ^3
\ \ \  {\rm G \; cm^3} \, ,
\label{EQN6}
\end{equation}
\

\noindent
where $R_{\rm WD}$ is the WD radius, 
and $B_{\rm WD}$ is randomly picked
from the probability density function
associated with the best-fitting gaussian.
As 
$B_{\rm WD}$ we treat $\mu_{\rm WD}$
in our simulations as being constant, 
for a particular polar 
throughout its evolution.

\begin{figure}
   \begin{center}
    \includegraphics[width=0.99\linewidth]
    {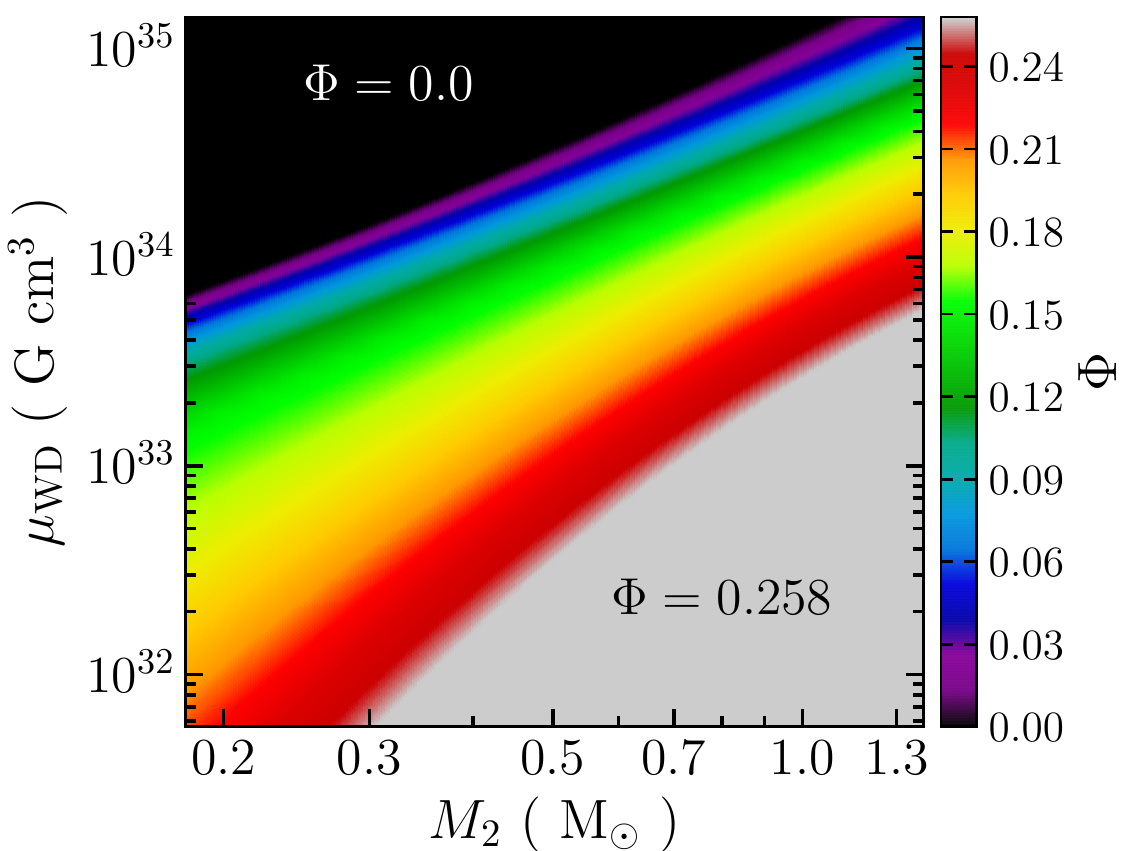}
    \end{center}
  \caption{The fraction of open field
  lines $(\Phi)$ as a function of the 
  donor mass $(M_2)$ and the WD magnetic
  moment $(\mu_{\rm WD})$. 
  The black area defines the region 
  where the magnetosphere  is completely  
  closed {${({\Phi=0.0})}$}, which provides that MB  
  is completely suppressed,  and  only
  GR plays a role in the polar evolution.
  The gray area corresponds to the non-magnetic
  case {${({\Phi=0.258})}$}.
  }
  \label{Fig02}
\end{figure}

In order to infer the value of $\Phi$,
we utilized fig.~3 in \ww, who showed 
that $\Phi$ is mainly a function of 
$\mu_{\rm WD}$ and $M_2$,
depending only very weakly on 
$M_{\rm WD}$.
In that figure, five values of $\Phi$ are 
given, from 0.0 to 0.2, in steps of 0.05.
We then linearly interpolate/extrapolate 
these curves to find the value of $\Phi$ 
for each polar in our simulations. 
The procedure is illustrated in 
Fig.~\ref{Fig02}, which shows the 
distribution of $\Phi$ in the plane 
$\mu_{\rm WD}$ versus 
$M_2$.
If a given polar in that
plane belongs to the region where
$\Phi=0.00$ 
(black area in the figure), 
MB is assumed to be absent,
and the evolution is driven by GR only.
Otherwise, MB is still present, and has 
its strength reduced as $\mu_{\rm WD}$ 
becomes larger or $M_2$ becomes smaller.
In the region where $\Phi=0.258$
(gray area in the figure), 
the evolution of the systems is assumed to be that of 
a non-magnetic CV.
Unlike $B_{\rm WD}$ and $\mu_{\rm WD}$,
$\Phi$ is not constant
for a given polar. Indeed, as shown 
in Fig.~\ref{Fig02}, $\Phi$ depends
on both $\mu_{\rm WD}$ and $M_2$.
Since $M_2$ decreases with time, 
as a result of mass transfer during 
CV evolution,
polars evolve through Fig.~\ref{Fig02}
from right to left at constant
$\mu_{\rm WD}$.
Hence, for each time-step,
$\Phi$ and the properties 
that depend on it
(Eqs. 
\ref{EQN1},
\ref{EQN3},
\ref{EQN4}
and
\ref{EQN5}
)
need to be updated.
A comparison between our approach
and \ww~is provided in 
Appendix~\ref{compWW02}.

%%%%%%%%%%%%%%%%
%%%%%%%%%%%%%%%%
%%%%%%%%%%%%%%%%
% NEW SECTION
%%%%%%%%%%%%%%%%
%%%%%%%%%%%%%%%%
%%%%%%%%%%%%%%%%

\section{Binary Population Modelling}
\label{bps}

To carry out our simulations,
we utilized the Binary Star Evolution 
(\bse) code 
\citep{Hurley_2000,Hurley_2002},
updated\,\footnote{
\href{http://www.ifa.uv.cl/bse}{\texttt{http://www.ifa.uv.cl/bse}}
}  by
\citet{Belloni_2018b}.
\bse~models AML mechanisms, such as
GR and MB, and
mass transfer occurs if either star 
fills its Roche lobe and 
may proceed on a nuclear, thermal, 
or dynamical time-scale.
The current version of the \bse~code 
includes state-of-the-art 
prescriptions for CV evolution, which 
allows accurate modelling of 
these interacting binaries. 
More details can be found in 
\citet{Belloni_2018b}.

Our modelling here
follows a similar
approach to that in \citet{Belloni_2018b}.
We carried out binary population synthesis 
using an initial population of 
$2\times10^8$ objects 
(single and binary stars)
assuming solar metallicity 
(i.e. $Z=0.02$).  
The properties of the binary stars are assumed to  
follow particular distributions.
First, the primary is picked from the
\citet{Kroupa_2001} initial 
mass function enforcing that
${1~{\rm M}_\odot~\leq~M_1~\leq~8~{\rm M}_\odot}$.
The secondary is then drawn from
a uniform mass ratio distribution, 
ensuring that $M_2 \leq M_1$, and 
that 
{${M_2\geq0.07}$~\Msun}.
The semimajor axis 
${(10^{0.5}~{\rm R}_\odot~\leq~a~\leq~10^{4.5}~{\rm R}_\odot)}$
and 
the eccentricity
${(0~\leq~e~\leq~1)}$
are assumed to follow a 
log-uniform 
and a thermal
distribution,
respectively.
Single stars are 
generated from the same initial mass function
as the primaries in binaries.

Regarding common-envelope evolution, 
we adopted a low efficiency, i.e. 
we assumed that 25 per cent of the 
dissipated binary orbital energy 
is used to expel the common envelope.  
The binding energy parameter 
was calculated according to 
\citet{Claeys_2014}
assuming that 
contributions 
from recombination energy 
are negligible
\citep[e.g.][]
{Zorotovic_2010,
Toonen_2013,
Camacho_2014,
Cojocaru_2017,
Belloni_2019a}.

As the \bse~code 
cannot handle thermal time-scale mass transfer,
we do not consider CVs 
emerging from this channel. 
This appears to be a reasonable
assumption as observations
show that only $\approx5$\,per cent of all CVs 
originate from this phase
\citep{Pala_2019b}.

Finally, the consequential angular 
momentum prescription adopted 
here is the one postulated by 
\citet{Schreiber_2016}, which is 
currently the only model that can 
explain 
crucial observations related to CV evolution,
such as the 
space density, 
the orbital period and WD mass distributions
\citep{Schreiber_2016,Belloni_2018b,McAllister_2019}.
Furthermore, 
it also provides an 
explanation for
the properties of detached CVs crossing
the orbital period gap
\citep{Zorotovic_2016},
the existence of single He-core WDs 
\citep{Zorotovic_2017},
and the mass density of CVs in 
globular clusters
\citep{Belloni_2019a}.

As in \citet{Goliasch_2015} and 
\citet{Belloni_2018b}, we assumed 
that the initial mass function 
is constant in time 
and that the binary fraction is 
50 per cent 
\citep[consistent with the binary 
fraction of WD primary progenitors, 
see][]{Patience_2002}.
Additionally, 
during the Galactic disc life-time, 
which is here assumed to be 
$\approx$~10~Gyr  
\citep{Kilic_2017},
we adopted a constant 
star formation rate 
\citep[e.g.][]
{Weidner_2004,
Kroupa_2013,
Recchi_2015,
Schulz_2015}. 
This implies that the birth-time 
distributions of 
both binary and single stars 
 are uniform.
The generated populations of single and binary stars are then evolved with the \bse~code from the birth time until the assumed Galactic age of 10 Gyr.

The CV space density is calculated 
following the scheme presented 
in 
\citet[][see their sections 2.2.3 
and 2.2.4]{Goliasch_2015}. 
As we generate both single stars and binaries,  
we can normalize the results of our 
population synthesis such that the number
of single WDs corresponds to a specific 
birth rate of WDs in the Galactic disc.
We adopt a WD space density of 
$\sim5\times10^{-3}$~pc$^{-3}$
\citep{Holberg_2008,Holberg_2016,
Esteban_2018,Hollands_2018} 
and a WD birth rate of
$\sim10^{-12}$~pc$^{-3}$~yr$^{-1}$
\citep{Vennes_1997,Holberg_2016},
which implies a WD formation rate 
of $\sim0.4$~yr$^{-1}$, 
and provides a total number of 
$\sim4\times10^9$~WDs in the disc. 
In order to derive the absolute number 
of systems that should be present
in the Galactic disc,
we similarly scaled the total number 
of CVs.
The space density is then computed by 
assuming a Galactic volume of 
$5\times10^{11}$~pc$^3$
\citep[e.g.][]{Toonen_2017}.

Finally, in our simulations, 
we assume that the fraction of 
polars relative to the entire CV population 
is $\simeq28$ per cent, which is the
measured fraction of polars
in the solar neighbourhood
\citep{Pala_2019b}. 
To simulate polar and non-magnetic CV evolution simultaneously we defined at
the onset of mass transfer whether the CV is non-magnetic or polar
assuming a probability of 28 per cent of
being a polar. The field strengths for the 
selected polars were than randomly drawn from the 
observed field strength distribution described in Section~\ref{obsBwd}.

%%%%%%%%%%%%%%%%
%%%%%%%%%%%%%%%%
%%%%%%%%%%%%%%%%
% NEW SECTION
%%%%%%%%%%%%%%%%
%%%%%%%%%%%%%%%%
%%%%%%%%%%%%%%%%
\section{Comparing model predictions and observations}
\label{sample}

In order to compare our predicted 
orbital period distributions
with observations,
we searched for all CVs
with accurate orbital period 
determinations 
from the 
{\it Sloan Digital Sky Survey}  (SDSS).
We found 199 systems, which are 
listed in Table~\ref{TabOBS}.
SDSS provides both photometric and 
spectroscopic data,
and allows the unambiguous 
identification of CVs 
from hydrogen
and helium emission lines.
In addition, its deep magnitude limit 
allows the detection
of intrinsically faint systems
\citep{Gansicke_2009},
characterized by low accretion rates, 
even at several scale heights above 
the Galactic plane.
This makes this sample
one of the largest homogeneous
CV samples available, as its broad 
colour selection 
range is superior to any previous 
surveys.

Recently, \citet{Pala_2019b} investigated
the intrinsic Galactic CV population in
a volume-limited sample of CVs, which is 
about 75 per cent complete.
This sample was built using accurate 
parallaxes provided by the 
\textit{Gaia} Data Release 2 parallax 
measurements and is composed of 42 CVs 
within 150\,pc, including two newly 
discovered systems.
The derived CV space density
is clearly better than
all previous 
estimates with high accuracy and small 
errors.
In addition, they measured the
intrinsic fraction of CVs above and below
the orbital period gap as well as the 
fraction of magnetic CVs.
As the 150\,pc sample is very small, 
we compare the predicted orbital period 
distributions with  
the homogeneous and large 
SDSS sample. We consider the less biased, but small, 150\,pc sample 
for the comparison between predicted and observed space densities
and fractions of period bouncers. 

To confront predicted 
mass transfer rates with observations, 
we compare predicted and observed effective temperatures 
of the WDs in CVs. The effective temperatures of these WDs are known 
to be reasonable tracers of the mean mass transfer rates,
as they depend on compressional heating 
caused by the accreted matter 
\citep[e.g.][]{Townsley_2003,Townsley_2004}.
When the mass transfer rate is rather low,
for dwarf novae in quiescence, or 
polars and nova-likes in the low state,
the emission is dominated by the WD 
surface, 
and the measurement 
of the WD effective temperature
is possible.
For the comparison of observed and predicted effective temperatures 
we used the 
samples from 
\citet[][and references therein, 
for non-magnetic CVs]
{palaetal_2017} and 
\citet[][and references therein, 
for polars]{TG09}.
These authors provide lists of CVs 
with reliable WD effective temperature
measurements 
and they contain in total
59 non-magnetic CVs and 13 polars.
We excluded
from the observational sample the six 
non-magnetic systems with evolved donors 
reported by \citet{palaetal_2017}, namely
V485 Cen, 
QZ Ser, 
SDSS J013701.06$-$091234.8, 
SDSS J001153.08$-$064739.2, 
BD Pav, and 
HS 0218$+$3229,
since we do not account for this
evolutionary channel in 
our simulations.

%%%%%%%%%%%%%%%%
%%%%%%%%%%%%%%%%
%%%%%%%%%%%%%%%%
% NEW SECTION 
%%%%%%%%%%%%%%%%
%%%%%%%%%%%%%%%%
%%%%%%%%%%%%%%%%
\section{Results}
\label{results}

\begin{figure*}
   \begin{center}
    \includegraphics[width=0.99\linewidth]
    {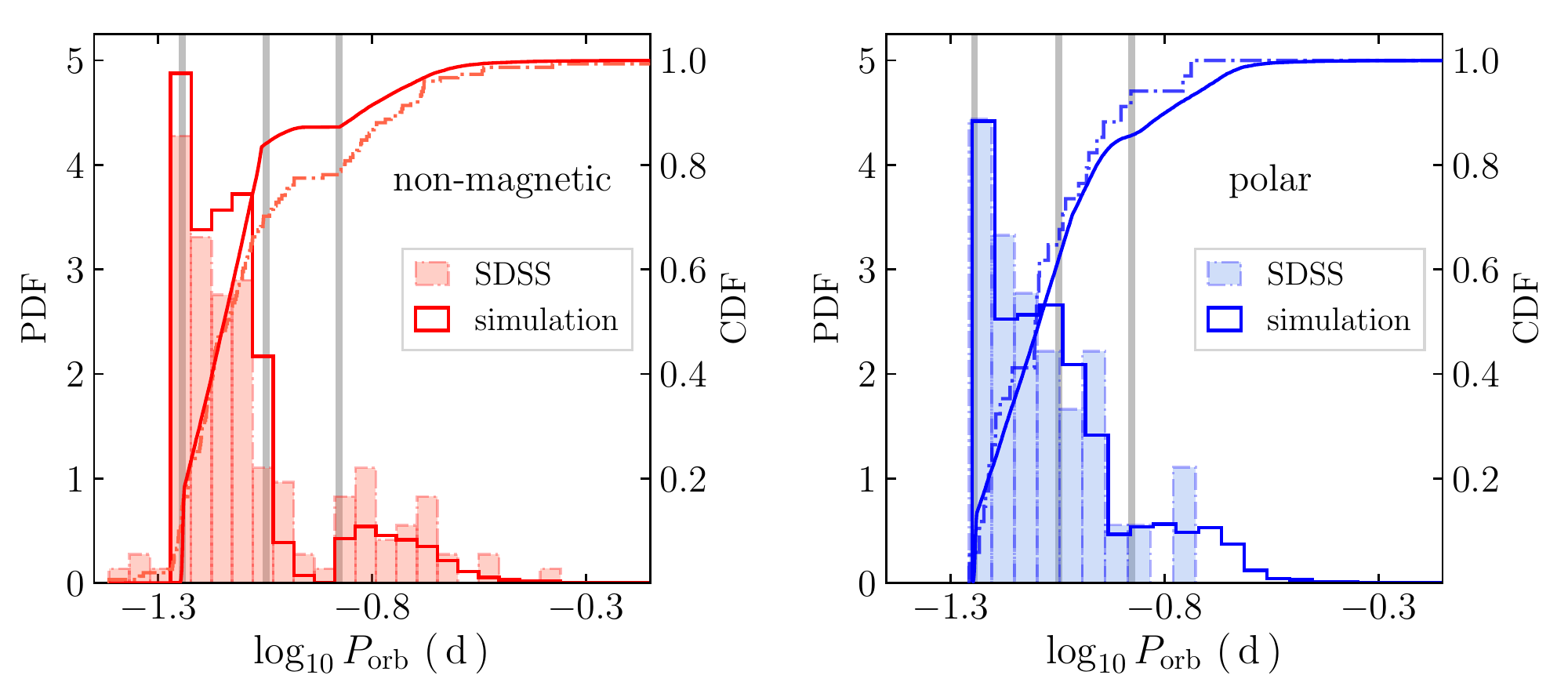}
    \end{center}
  \caption{Comparison between observed 
  and predicted orbital period 
  $(P_{\rm orb})$ distributions 
  for both non-magnetic CVs (left-hand panel) 
  and polars (right-hand panel).
  Observed measurements for 
  non-magnetic CVs and polars are 
  from SDSS data (Table~\ref{TabOBS}).
  Vertical lines in the period 
  distribution are the observational
  location of the period minimum 
  \citep{Gansicke_2009} and 
  gap edges \citep{Knigge_2006}.
  Notice that distributions for both 
  non-magnetic CVs and polars
  strongly differ from each other.
  See text for more details.
  }
  \label{Fig04}
\end{figure*}

\begin{figure*}
   \begin{center}
    \includegraphics[width=0.99\linewidth]
    {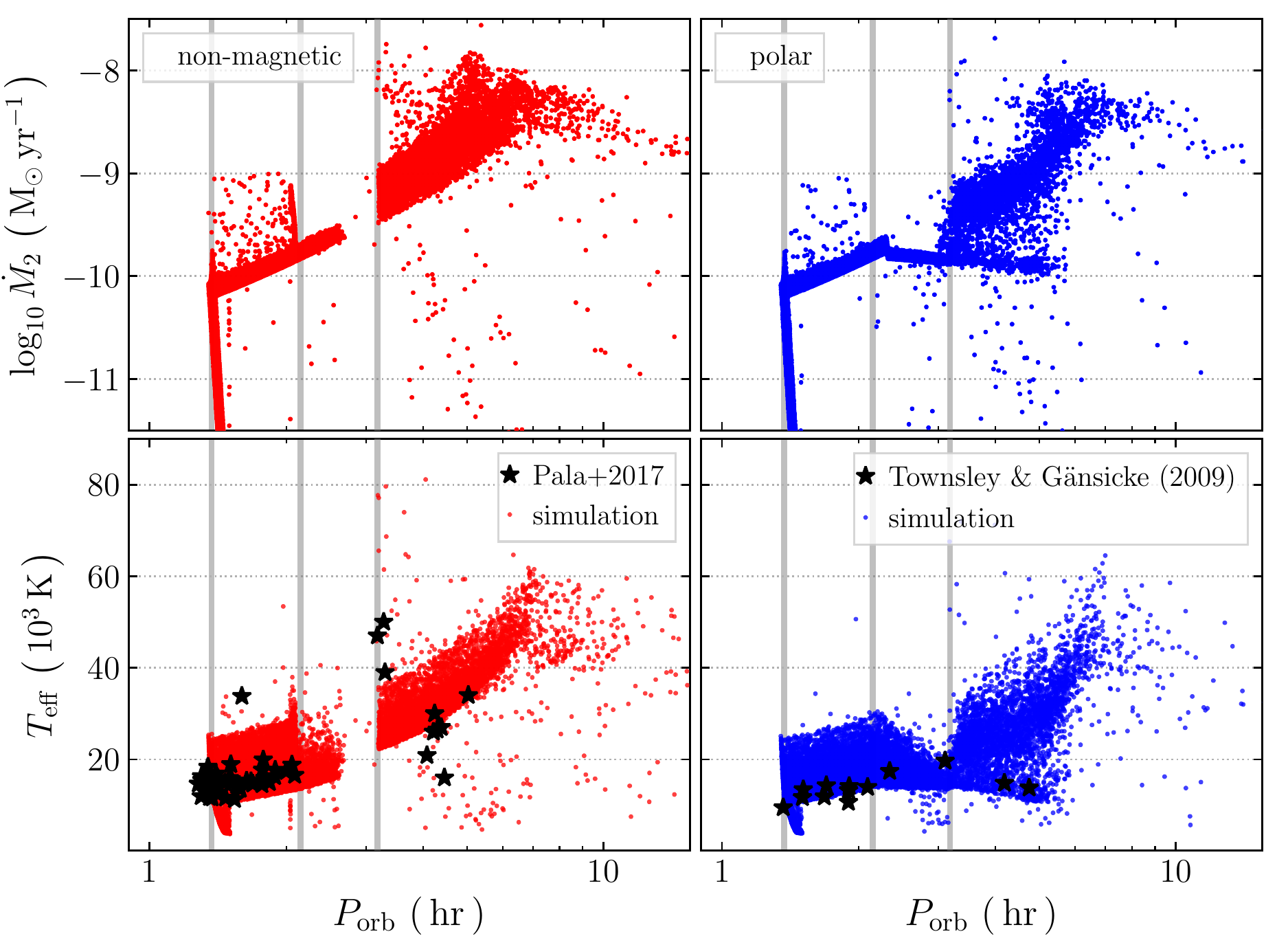}
    \end{center}
  \caption{Predicted mass transfer 
  rate ($\dot{M}_2$, top panels) and 
  WD effective temperature
  ($T_{\rm eff}$, bottom panels) against
  the orbital period ($P_{\rm orb}$).
  In the left-hand panels non-magnetic 
  CVs are shown, and
  in the right-hand panels polars 
  are depicted.
  Observed values related to 
  $T_{\rm eff}$ are shown as
  black stars and are from  
  \citet[][and references therein, 
  for non-magnetic CVs]
  {palaetal_2017} 
  and 
  \citet[][and references therein, 
  for polars]
  {TG09}.
  Vertical lines in the period 
  distribution are the observational
  location of the period minimum 
  \citep{Gansicke_2009} and 
  gap edges \citep{Knigge_2006}.
  Notice that mass transfer rates of
  polars above the orbital period
  gap are in general smaller than 
  those of non-magnetic CVs, which
  consecutively results in smaller 
  $T_{\rm eff}$ amongst polars.
  }
  \label{Fig05}
\end{figure*}

We concentrate the analysis
of our population synthesis 
outcomes on three very important CV
observables, namely 
the orbital period distribution,
the mass transfer rate distribution
(or alternatively 
the WD effective temperature
distribution) and
the space density.
While comparing 
the orbital period
distributions, we will 
consider only systems with donor masses
greater than {$0.05$~\Msun}, 
i.e. we neglect period bouncers. 
The reason is twofold. 
First, 
the mass-radius relation 
for CV donors having {$M_2<0.05$~\Msun}
is not reliable
\citep{Knigge_2011_OK}.
Secondly, 
period bouncers are extremely rare
in observed samples
\citep[e.g.][]
{Santisteban_2018,Pala_2019b}.
The observed sample of period bouncers is therefore most likely 
much more biased and incomplete than the 
sample of brighter CVs.

%%%%%%%%%%%%%%%%
% NEW SUBSECTION
%%%%%%%%%%%%%%%%
\subsection{Orbital period distribution}
\label{pdist}

Starting with
the observed distributions
({NON-MAG} and {POLAR} systems in 
Table~\ref{TabOBS}),
we see a gap in the orbital
period distribution of non-magnetic CVs,
while the same feature is absent in polars.
Indeed, the fraction of non-magnetic CVs
inside the gap 
(i.e. 2~h~$<$~$P_{\rm orb}$~$<$~3~h) 
is $\approx8$ per cent,
while for polars this fraction
is much larger, being 
$\approx25$ per cent.
We also see a larger fraction of 
non-magnetic CVs above the gap 
in comparison to polars.
In particular, the fraction of polars with
periods longer than $\approx3$~h 
is only $\approx9$ per cent, 
while $\approx21$ per cent of 
non-magnetic CVs are located above the gap.
These two features put together
suggest that both observed distributions
are intrinsically different, even though
we still rely on relatively small-number
statistics for such a claim.

In agreement with observations, 
our population synthesis shows significant differences in the predicted orbital period distributions.
On the left-hand panel of Fig.~\ref{Fig04}, 
we compare the predicted orbital 
period distribution of non-magnetic 
CVs with the observed one.
Notice that the results from our simulation 
reasonably well reproduce the main
features of the
observed distribution, 
i.e. the orbital period gap between
$\approx2$ and $\approx3$~h, and the 
accumulation of systems close to the 
orbital period minimum ($\approx80$ min).
We can also reasonably well reproduce the
decreasing number of systems above the 
orbital period gap towards longer periods.
The orbital period distribution of polars 
(right-hand panel of Fig.~\ref{Fig04}) 
differs from 
the distribution of non-magnetic CVs.
In the polar case, 
instead of presenting an orbital
period gap, polars gradually fill the region
between $\approx2$ and $\approx3$~h, as a 
consequence of the reduction of AML through MB.
Indeed, for sufficiently high values
of $\mu_{\rm WD}$, MB becomes considerably less efficient, which
leads to smaller mass transfer rates 
above the orbital period gap and 
donors less bloated.
As a consequence, the contraction of the donor stars in these systems 
when MB is disrupted is less significant and 
the polar needs less time to become 
semidetached again.
The overall effect is a dilution of 
the gap, as a relatively large number 
of polars penetrates the region of the 
gap when mass transfer turns on again.
Concerning the orbital period minimum, 
we do not detect significant difference between polars and non-magnetic CVs. 
In both cases we detect a 
peak at $\approx80$ min.

In order to evaluate whether our results are 
consistent with observed distributions on 
statistical grounds, we applied 
two-sample Kolmogorov--Smirnov
tests. 
The null hypothesis of each test
is that {\it both predicted and
observed distributions stem from 
the same parent population}.
Comparing 
%now 
predicted and 
observed distributions
in the non-magnetic case, 
the $p$-value is 0.001, 
which indicates that our predicted 
distribution differs from the 
observed one, even though we can 
clearly reproduce key features of 
the observed distribution.
The reason for that is likely an 
observational bias against the detection 
of short-period systems.
Regarding polars, the $p$-value 
is
{${\approx0.53}$}.
This provides strong support to 
the null hypothesis that
both distributions stem from the 
same parent distribution.
Consequently, this suggests that the 
reduced MB prescription represents an appropriate   
model to explain polar
evolution.

%%%%%%%%%%%%%%%%
% NEW SUBSECTION
%%%%%%%%%%%%%%%%
\subsection{Mass transfer rates and WD effective temperatures}
\label{twd}

A direct consequence of the 
reduced MB model in polars is 
the reduction of their mass transfer
rates above the orbital period gap, 
as well as the
slow-down of the evolution.
In the top row of Fig.~\ref{Fig05} we 
compare predicted
mass transfer rates 
$(\dot{M}_2)$
against 
the orbital period of 
non-magnetic CVs (left-hand panel) and 
polars (right-hand panel).
Note that, as expected, both classes 
have similar mass transfer rates
below the orbital period gap 
{$(\sim10^{-10}$\,\Msun\,yr$^{-1})$}. 
However, above the gap, 
most non-magnetic CVs 
have mass transfer rates between 
$10^{-9}$ and $10^{-8}$~\Msun~yr$^{-1}$, 
while most polars have
{${\dot{M}_2\lesssim10^{-9}}$~\Msun~yr$^{-1}$}.
This is a difference of
one to two orders of magnitude.

A way of testing whether such 
predictions are consistent with 
observations is by means of the
WD effective temperature. 
The WD effective temperatures in CVs trace the mean mass transfer 
rate as the WD is heated 
%compressional
%heating, i.e. 
by the energy released 
by fluid elements
as they are compressed by 
further accretion 
\citep{Townsley_2003,Townsley_2004}.
The WD effective temperature 
($T_{\rm eff}$) due
to this compressional heating is 
given by 
\citep[][eq. 2]{TG09}

\begin{equation}
\frac{T_{\rm eff}}{{\rm K}} \ = \ 1.7\times10^4 \; \left( \, \frac{\langle \, \dot{M}_2 \, \rangle}{10^{-10} \, {\rm M}_\odot \, {\rm yr}^{-1}} \, \right)^{0.25} \; \left( \, \frac{M_{\rm WD}}{0.9 \, {\rm M}_\odot} \, \right), \label{compheat}
\end{equation}
\

\noindent
where $\langle \dot{M}_2 \rangle$ 
and $M_{\rm WD}$ are 
the average mass transfer rate 
and the WD mass, respectively.
However, this approach is only correct if the 
WD cooling temperature is smaller than 
the temperature derived from 
compressional heating. 
We therefore also determined 
the WD cooling temperature
as in 
\citet{Zorotovic_2017},
i.e. via interpolation of DA 
(pure hydrogen atmosphere) 
WD evolutionary  models by 
\citet{Althaus_1997}, 
for helium-core WDs, and by 
\citet{Fontaine_2001}, 
for carbon/oxygen-core WDs.
Having calculated both the cooling and 
the compressional heating temperatures, 
we took the higher 
temperature for each WD produced by our 
population model.

In the bottom row of 
Fig.~\ref{Fig05}, we compare 
predicted and observed
$T_{\rm eff}$ of non-magnetic CVs 
(left-hand panel) and 
polars (right-hand panel).
Observed values are from 
\citet[][and references therein]{palaetal_2017} 
and 
\citet[][and references therein]{TG09}.
As in the case of 
the mass transfer rate, below the gap
both CV types have similar 
$T_{\rm eff}$ in the range of
${\sim10\,000-20\,000}$~K,
and are in general in agreement
with the observations.
However, 
the predicted $T_{\rm eff}$ above 
the gap drastically differ 
from each other. 
As the mass transfer rates 
are smaller for polars in this 
period range, so are the values of 
$T_{\rm eff}$ (Eq. \ref{compheat}).
While the temperatures for 
non-magnetic CVs are usually 
greater than $\sim30\,000$ K, 
those of polars are in general 
smaller than that.

The predicted and observed 
$T_{\rm eff}$ of polars
seem to agree 
with each other, even though this 
claim is weaker for
systems above the orbital 
period gap, due to
small-number statistics.
Regarding non-magnetic CVs,
our predicted $T_{\rm eff}$
provides reasonable results 
below the orbital period gap
as most systems
fall in the predicted range with
the exception of one system. 
This CV (SDSS J153817.35$+$512338.0)
could be either a young CV or recently had 
a nova explosion \citep{palaetal_2017}.
However, above the
gap, only two systems (out of 10) 
are consistent with
predicted values. The remaining 
systems are either above (nova-likes) 
or below (dwarf novae) the range
of predicted $T_{\rm eff}$ which represents a striking disagreement 
between observations and theory. 
In any event, our results concerning polars
are encouraging and, together with the
already-mentioned orbital period 
distribution, provide strong  
support for the reduced MB model.

%%%%%%%%%%%%%%%%
% NEW SUBSECTION
%%%%%%%%%%%%%%%%
\subsection{Space density}
\label{space}

The last aspect we will address is 
the space density.
The numbers of polars and non-magnetic
CVs produced in our simulations
are provided in Table~\ref{Tab01}, 
together with the space density,
computed as described in Section 
\ref{bps}.
By construction, 
the predicted fraction of polars 
($\approx25$ per cent)
matches that found in observations, 
which is $\approx28$ per cent
\citep{Pala_2019b}.
The predicted space densities 
for all CVs,
non-magnetic CVs and
polars are 
${\approx2.2^{+2.2}_{-1.1}\times10^{-5}}$,
${\approx1.6^{+1.6}_{-0.8}\times10^{-5}}$ and
${\approx5.2^{+5.2}_{-2.6}\times10^{-6}}$~pc$^{-3}$,
respectively.
When removing the period bouncers,
those 
space densities are
${\approx4.5^{+4.5}_{-2.3}\times10^{-6}}$,
${\approx3.0^{+3.0}_{-1.5}\times10^{-6}}$, and
${\approx1.5^{+1.5}_{-0.8}\times10^{-6}}$~pc$^{-3}$,
respectively.
Due to uncertainties in 
the initial binary populations 
these space densities might be affected
by a factor of $2$.

\begin{table*}
\centering
\caption{Number and space density 
of non-magnetic CVs and polars 
in our simulations. 
The column labelled 
{\emph{Model}} 
presents results from 
our simulations, while 
the column labelled {\emph{Absolute}} 
shows the numbers scaled 
according to the birth rate of 
single WDs.
Finally, the space density ($\rho$) 
was calculated with the absolute 
numbers, assuming a Galactic 
volume of $5\times10^{11}$ pc$^3$. 
The errors are estimated from
uncertainties in the initial 
binary population.
In the last four columns we provide the 
fractions of long-period, short-period
and gap CVs, and period bouncers.
Finally, in the last column, we provide
the space density when period bouncers
are excluded.
}
\label{Tab01}
\begin{adjustbox}{max width=\linewidth}
\noindent
\begin{threeparttable}
\noindent
\begin{tabular}{l|c|ccc|cccc|c}
\hline
\hline
CV type & &
Model  & 
Absolute & 
$\rho$  &
$f_{\rm long}$ 	&
$f_{\rm short}$  	&
$f_{\rm gap}$   	&
$f_{\rm bouncers}$  	& 
$\rho$ (without bouncers) \\ 
\multicolumn{3}{c}{} & 
$\left(~10^6~\right)$ &
$\left(~10^{-6}~{\rm pc}^{-3}~\right)$ & 
(per~cent)	&
(per~cent)	&
(per~cent)	&
(per~cent)	& 
$\left(~10^{-6}~{\rm pc}^{-3}~\right)$ \\
\hline
all CVs & & 
$324509$ & $10.732$ & ${21.5^{+21.5}_{-11.3}}$
& 
$2.9$ &  $11.2$ &  $4.1$ &  $81.8$ 
&
${4.5^{+4.5}_{-2.3}}$ \\ 
\hline
non-magnetic & & 
$245393$ & $8.116$ & ${16.2^{+16.2}_{-8.1}}$
& 
$2.4$ &  $11.2$ &  $2.4$ &  $84.0$ 
&
${3.0^{+3.0}_{-1.5}}$ \\ 
\hline
polar & & 
$79116$ & $2.616$ & ${5.2^{+5.2}_{-2.6}}$
& 
$4.3$ & $11.3$ &  $9.3$ &  $75.1$ 
&
${1.5^{+1.5}_{-0.8}}$ \\ 
\hline
\hline 
\end{tabular}
\end{threeparttable}
\end{adjustbox}
\end{table*}

Predicted space densities are in reasonable
agreement with observations, within the errors.
For example, using 20 non-magnetic CVs, 
\citet{Pretorius_2012} 
determined a space density of
${4^{+6}_{-2}\times10^{-6}}$~pc$^{-3}$,
from the \textit{ROSAT} Bright Survey 
and the \textit{ROSAT} North Ecliptic Pole 
survey, which are supposedly
complete X-ray flux-limited
surveys.
\citet{Schreiber_2003} derived a lower 
limit of ${\sim10^{-5}}$~pc$^{-3}$ 
for the space density of
post-common-envelope binaries that
are CV progenitors.
\citet{Santisteban_2018} estimated 
an upper limit for the period bouncer 
space density of
${\lesssim2\times10^{-5}}$~pc$^{-3}$
using SDSS Stripe 82 data. 
\citet{Schwope_2018},
who took into account recent distance
measurements from the \textit{Gaia} satellite
in previous determinations using 
\textit{ROSAT} surveys,
measured a space density 
of non-magnetic CVs to be 
${\sim10^{-6}-10^{-5}}$~pc$^{-3}$,
depending on the assumed scale height
and survey.
Regarding polars, \citet{Pretorius_2013}
used 24 systems from the X-ray flux-limited 
\textit{ROSAT} Bright Survey sample 
and estimated a space density of
${9.8^{+5.4}_{-3.1}\times10^{-7}}$~pc$^{-3}$,
provided that their high-state duty cycle 
are 0.5 and that they are below the 
survey detection limit 
during their low states.

Most recently, \citet{Pala_2019b}
determined space densities with
unprecedented small uncertainties.
Their measured values are 
${4.8^{+0.6}_{-0.9}\times10^{-6}}$
and 
${1.2^{+0.4}_{-0.5}\times10^{-6}}$~pc$^{-3}$,
for all and magnetic CVs, respectively,
assuming a scale height of 280 pc.
These space densities are the most reliable 
ones derived from observations so far
and are smaller by a factor
of $2-6$ than our predictions. 
However, if we exclude period bouncers
from both predicted and observed samples 
(2 out of 42 systems in the 150\,pc sample
are potentially period bouncers), 
the predicted and observed  
space densities are in very good
agreement.

We also show in Table~\ref{Tab01} the relative 
fractions of simulated systems in different period ranges.
It is clear from the table that the fraction 
of polars inside the gap is much
larger than that of non-magnetic CVs 
(about four times). In addition, the fraction
of polars above the gap is also larger 
(about two times). Moreover, there is a reduction
of about 10 per cent of period bouncers in 
polars, and both have similar fractions
of systems between the period minimum 
and the lower edge of the gap.
The different fractions for non-magnetic 
and polar CVs, are caused by different 
characteristic evolutionary time-scales.
Indeed, in the reduced MB model, polars 
are less affected
by MB (several only by GR), due to the strong
influence of their high WD magnetic field on MB.
On the other hand, non-magnetic CV 
evolution above the gap is driven by AML 
due to full MB, which is around one order 
of magnitude stronger than GR, and greater than
any reduced MB. This results in a faster
evolution for non-magnetic systems, in 
comparison with polars. Therefore, 
the relative number of polars above
the gap is larger than for non-magnetic 
CVs. For the same reason, less polars manage
to become period bouncers.

Excluding period bouncers (which either 
seem to be difficult to find or are over predicted by our model),
the predicted fraction of all CVs above the gap is $\approx16$\,per cent.  
This prediction is in excellent agreement 
with the observations.
\citet{Pala_2019b} found
that {${\approx17.5}$}\,per cent
of all CVs are above the orbital
period gap, when their $\approx5$\,per cent 
period bouncer candidates are removed
from the analysis.

%%%%%%%%%%%%%%%%
%%%%%%%%%%%%%%%%
%%%%%%%%%%%%%%%%
% NEW SECTION
%%%%%%%%%%%%%%%%
%%%%%%%%%%%%%%%%
%%%%%%%%%%%%%%%%
\section{Discussion}
\label{discussion}

We have shown in Section~\ref{results}
that the space densities and 
orbital period distribution of 
both non-magnetic CVs and polars agree
well with observations if period bouncers
are excluded from the analysis.
In addition, the mass transfer rates of 
polars also seem to agree with the 
observations while those predicted for non-magnetic CVs cannot reproduce the mass transfer rates derived from observations for systems above the gap.
Here, we will discuss potential caveats 
in both simulations and observations.

%%%%%%%%%%%%%%%%
%%%%%%%%%%%%%%%%
%%%%%%%%%%%%%%%%
% NEW SECTION
%%%%%%%%%%%%%%%%
%%%%%%%%%%%%%%%%
%%%%%%%%%%%%%%%%
\subsection{Is the SDSS CV 
sample biased against 
short-period systems?}
\label{sdssbias}

We showed in Section \ref{pdist}
that the reduced MB model proposed
by \lww~is capable of providing
a satisfactory explanation for differences
in the observed orbital period
distributions of non-magnetic CVs 
and polars.
Even though our predicted orbital period 
distribution for non-magnetic CVs exhibits
the key features of the observed distribution,
we found statistical evidence
supporting the hypothesis that they do not
stem from the same parent population.
Such a result is likely due to observational
biases involved in the SDSS CV sample.

\begin{figure}
   \begin{center}
    \includegraphics[width=0.99\linewidth]
    {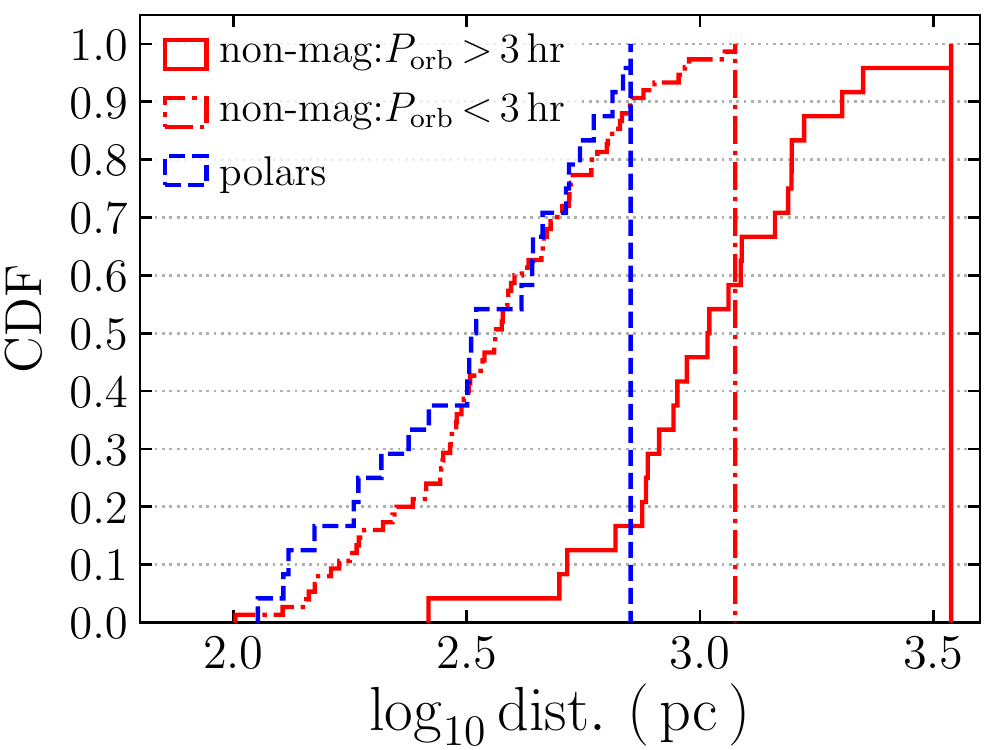}
    \end{center}
  \caption{Cumulative distribution of 
  SDSS CV distances
  whose measurements are reliable 
  (errors smaller than
  30 per cent).
  The red solid line is for 
  non-magnetic CVs with periods 
  longer than 3~h, 
  the red dot-dashed line is for
  non-magnetic CVs with periods 
  shorter than 3~h
  and 
  the blue dashed line is for 
  polars.
  Notice that the distances to
  polars are on average
  smaller, followed by 
  short-period non-magnetic
  CVs. Non-magnetic CVs 
  above the orbital period
  gap have on average the 
  largest distances among
  SDSS CVs.
  }
  \label{Fig06}
\end{figure}

As discussed in
\citet{Gansicke_2009},
SDSS covers high Galactic 
latitudes ($|b|>30^{\rm o}$) and
could identify long-period CVs 
out to a distance of 
$\gtrsim10^4$~pc, while 
WD-dominated CVs are found only out to a 
distance of 
$\sim300$~pc.
This difference in distance 
illustrates the bias towards 
long-period CVs in 
comparison to 
CVs with shorter periods.
Further evidence is provided in 
Fig.~\ref{Fig06}, which 
shows the cumulative distributions 
of SDSS CV distances
for CVs whose distance 
measurements have errors
smaller than $30$ per cent.
Approximately $3/4$ of the 
CVs in the SDSS sample 
fulfil this criterion.
The average distance for 
non-magnetic CVs with
periods longer than 3~h 
is $1\,206\pm666$~pc,
while for non-magnetic CVs 
with periods shorter
than 2~h, it is $422\pm231$~pc.
Consequently, a better agreement between 
predicted and observed
distributions would likely be 
achieved if more short-period
non-magnetic CVs could be detected,
whose reduced relative number is 
expected to be a consequence of the 
limiting magnitude of the SDSS.

Provided that both non-magnetic 
CV and polar samples
come from SDSS, one might 
ask why the bias previously
discussed does not influence 
the comparisons in the 
case of polars, in which 
we found that both predicted
and observed distributions are 
consistent and likely stem from 
the same parent population.
Figure~\ref{Fig06} 
shows the cumulative 
distribution of polar distances, 
for those with reliable distance 
measurements (error smaller than
30 per cent).
On average, polars have 
distances of $369\pm183$~pc,
which implies that they 
are considerably closer than 
long-period non-magnetic CVs.
Due to reduced MB the mass transfer differences in polars
and its dependence on orbital period are much smaller
than in non-magnetic CVs.   
Therefore the distance/magnitude bias is
weaker than
in non-magnetic CVs and 
the SDSS polar 
sample is more 
representative of the true polar 
population in the Galaxy.
In order to verify this we performed a
two-sample Kolmogorov--Smirnov test 
taking into account both the 
SDSS polar sample (34 systems) 
and 
the 150\,pc polar sample (12 systems).
The $p$-value is $\approx0.151$,
which does not allow us to reject 
the null hypothesis that indeed 
both distributions stem 
from the same parent population.

Regarding the relative incidence of non-magnetic
CVs above and below the orbital period gap, 
we note that we predict similar 
fractions of systems below 
the orbital period gap than found in the 
SDSS CV sample.
The fraction of non-magnetic CVs 
above the orbital period gap in the SDSS 
sample is $21\pm3$~per cent,
and $79\pm3$ per cent 
are either in the orbital period gap or below.
The same fractions
in the predicted population (excluding
period bouncers) are
$15$ and $85$~per cent, 
respectively.
In the 150\,pc sample 
\citep{Pala_2019b},
the same fractions for non-magnetic CVs 
are $18\pm7$ and $82\pm7$ per cent, 
which are very close to the predicted
values.

We therefore conclude that the SDSS sample is clearly still biased towards long-period non-magnetic systems and that this is likely the reason for the disagreement 
in the left-hand panel of 
Fig.~\ref{Fig04}.

%%%%%%%%%%%%%%%%
%%%%%%%%%%%%%%%%
%%%%%%%%%%%%%%%%
% NEW SECTION
%%%%%%%%%%%%%%%%
%%%%%%%%%%%%%%%%
%%%%%%%%%%%%%%%%
\subsection{What problems are
still standing in CV evolution?}
\label{caveats}

In Section \ref{space},
we computed the space
densities for non-magnetic 
CVs and polars and they are 
generally in good agreement with
previous and rather crude observational 
estimates.
The by far most reliable measurements of CV space densities \citep{Pala_2019b}, however, 
are a factor of $2$--$6$ smaller than our predictions. This is entirely caused by the fact that our models predict $80$~per~cent of all CVs to be period bouncers while only $\simeq5$~per~cent of the CVs within 150\,pc seem to be period bouncers. If we compare the space densities of CVs prior to the period minimum, our predictions perfectly agree with the space densities derived from the 150\,pc sample. 

We see two possible explanations for this disagreement between predicted and observed number of period bouncers. 
First, the absence of period bouncers in the 150\,pc sample could mean that they do not exist in numbers as large as predicted by our evolutionary models. This suggests that CVs reach the period minimum (because the period distribution below the gap roughly agrees with the predictions) but that they have not had time to evolve past the period minimum. This would imply that
the time-scale for an initial binary to evolve into a CV would be significantly longer and the currently observed CVs would therefore be significantly older systems (the initial binary was born earlier). 
Alternatively, even the 150\,pc sample could miss a large number of existing period bouncers as their mass transfer rates can be extremely low and their outburst frequency can be extremely long. Dedicated deep surveys for period bouncers are the only possibility to solve this issue.

Regarding mass transfer rates and WD effective 
temperatures due to compressional
heating, we showed in Section~\ref{twd} that 
our results are in good agreement with
observed values of polars and non-magnetic
CVs below the orbital period gap.
However, as we can clearly see from 
Fig.~\ref{Fig05}, predicted values 
for non-magnetic CVs above the gap
drastically disagree with observations. 
This issue is very likely not caused by an observational bias, as some of the systems the model does not reproduce have higher and others have lower temperatures than predicted by our model. 

In general we see two possibilities to solve this puzzle. 
First, the measured WD effective temperatures could not represent a good proxy of the accretion rate. 
Indeed, the temperature of the WD is only sensitive to variations of the mass transfer rate on time-scales that are much shorter than those produced by the most likely 
types of long-term fluctuations
\citep[e.g. irradiation-driven cycles or nova-induced variations,][]{Knigge_2011_OK},
which could result in unreliable estimates
of the secular/average mass transfer rate for
some systems.
More importantly, in CVs that have undergone many nova eruptions,
the WD cooling could be affected by changes in the structure of its outer envelope
and for these systems the accretion rate could thus be overestimated 
by only considering compressional heating.

At first glance, this idea appears plausible because most 
novae are observed in the period range above the gap, especially between $3$ and $4$~h, exactly where the high-temperature WDs are found \citep{Tappertetal2017}. 
However, if the mass transfer rates were much lower than those derived from compressional heating, the cooling time-scale of WDs heated by nova eruptions \citep{Prialnik1986} would be about an order of magnitude shorter than the corresponding nova cycle. To have caught all three systems in the $3$--$4$~h period range in the relatively 
short post-nova phase therefore appears to be extremely unlikely. 
It seems more plausible to assume that the origin of the high temperatures are large mass transfer rates. 
As shown by \citet{Townsley_2005},
the accumulation of novae above the period gap is  perfectly consistent with the high mass transfer
rates simply because larger mass transfer rates lead to more frequent nova eruptions. While these high mass transfer rates could shorten the nova cycle and thus increase the probability to observe systems in the post-nova phase, the conclusion that the measured high temperatures imply large mass transfer rates would not be affected. We therefore conclude that most likely the mass transfer rates are indeed 
very high in the systems with hot WDs.

The second, and much more likely, possible solution for the observed disagreement is that the model is incomplete.
If the mass transfer rates are indeed as high as indicated by the WD temperatures, 
we are clearly missing a fundamental ingredient in our models of CV evolution above the gap. While the overall evolution predicted by our models is
roughly correct (as the period distribution 
reasonably well agrees with the observations), something seems to be
missing. 
Maybe the period dependence of AML through MB 
is significantly different from
what we assumed here and/or AML for systems above the orbital period gap might vary not only with the orbital period. Perhaps the problem of the mass transfer rates above the orbital period gap and the missing period bouncers are both related to our limited understanding of MB. It would certainly be interesting to test if revisions of MB can fix this problem while keeping the otherwise good agreement between theory and observation.

%%%%%%%%%%%%%%%%
%%%%%%%%%%%%%%%%
%%%%%%%%%%%%%%%%
% NEW SECTION
%%%%%%%%%%%%%%%%
%%%%%%%%%%%%%%%%
%%%%%%%%%%%%%%%%
\subsection{What is the origin of 
magnetic WDs?}

In this work we performed the first binary population synthesis for polars and obtained excellent agreement between theory and observations. 
In particular we investigated 
how the evolution is affected 
by strong WD magnetic fields. 
However, we assumed in the simulations
that the theoretical
distribution of WD magnetic fields matches the 
observed one and that the field strength remains 
constant throughout CV evolution
(Section \ref{obsBwd}).

However, this assumed distribution 
might not be representative of the 
intrinsic polar population in the Galaxy,
since we used the polars listed
in the review by \citet*{Ferrario_2015},
which is likely a biased sample
of polars.
This is because this sample is simply
a list of polars discovered so far with
determined properties, which suffers 
from different biases. 
Polars are strong X-ray emitters
and usually detected/discovered by 
high-energy missions such as 
\emph{ROSAT}, 
\emph{XMM--Newton}, 
and \emph{Swift},
and therefore the observed polar sample might be biased with respect to e.g. X-ray flux limits or restrictions in Galactic latitude and distance.
In addition,
given that the origin of the strong WD magnetic fields 
is still not understood, 
the field strength could eventually change over time as CVs evolve.

There are currently three main scenarios that account for the formation of magnetic fields in WDs that are applicable to close binaries.
In the fossil field scenario \citep[e.g.][]{Angel_1981}, magnetic Ap and Bp stars are the progenitors of magnetic WDs, provided the magnetic flux is conserved till the WD formation. However, \citet{Kawkaetal_2007} showed that the
birth rate of magnetic Ap and Bp stars is 
too small to explain the relatively large number of 
magnetic WDs.
Alternatively, \citet{Tout_2008} proposed that the origin of high magnetic fields in WDs is a magnetic dynamo acting during the common-envelope phase. In a parallel effort, we used the numerical code presented in this work to test this common-envelope dynamo scenario and find that its predictions do not agree with the observed properties of close magnetic WD binaries.
For example, the common-envelope dynamo scenario struggles to explain the fact that 
all observed pre-polars contain only old and cool WDs
\citep
[$T_{\rm eff}\lesssim10\,000$~K;~e.g.][]
{Reimers_1999,Reimers_2000,Schwope_2002a,Schmidt_2005,Schwope_2009,Parsons_2013}
while not a single detached magnetic CV progenitor system with a young hot WD has been found
\citep[see][for more details]{Belloni_2020}. 
As a third alternative for magnetic field generation in WDs, \citet{Isern_2017} argued that when the WD temperature is low enough, its interior crystallizes, which in turn 
allows the generation of
a magnetic field through a dynamo 
similar to the ones operating 
in either stars or planets. 
However, the field strengths predicted by the current version of this model are far lower than those observed in magnetic CVs. 

Thus, despite several scenarios being proposed, we currently lack a model that correctly reproduces the 
observations of magnetic WDs in binaries. 
One likely possibility is that not only one channel contributes to the production of magnetic WDs and that their contributions might be different for different types of systems.

%%%%%%%%%%%%%%%%
%%%%%%%%%%%%%%%%
%%%%%%%%%%%%%%%%
% NEW SECTION
%%%%%%%%%%%%%%%%
%%%%%%%%%%%%%%%%
%%%%%%%%%%%%%%%%
\section{Summary and Conclusions}

We performed population synthesis of CVs composed of highly magnetized WDs (i.e. polars) and non-magnetic WDs, and predicted period and mass transfer rate distributions, as well as space densities for both populations. The presented calculations are the first binary population models that properly include reduced magnetic braking in polars.

We found that the mass transfer rates for CVs below the orbital period gap are in agreement with the observations for both non-magnetic CVs and polars. Above the gap, the mass transfer rates derived for polars also agree with the observations, while those predicted for non-magnetic CVs drastically disagree. The latter implies that we most likely do not properly understand the strength and dependencies of angular momentum loss due to magnetic braking and a different prescription could potentially solve the discrepancy in the mass transfer rates of long-period non-magnetic CVs.

The predicted orbital period distribution for polars differs from that of non-magnetic CVs, and is consistent with the differences in the observed distributions. The period gap is absent in the predicted polar distribution due to the reduced magnetic braking. The predicted orbital period distribution for polars nicely agrees with the observed one, which provides strong support for the reduced magnetic braking hypothesis. For non-magnetic systems, we find reasonable agreement between the observed and predicted orbital period distributions, provided observational biases are taken into account.

Finally, we predicted space densities which are slightly larger (by a factor of $2-6$) than those derived from the 150\,pc sample. However, this disagreement can be entirely explained by the different fractions of period bouncers. While the observed sample includes only $\simeq5-10$ per cent period bouncers, the simulated sample is dominated by systems that already passed the period minimum ($\simeq80$ per cent are period bouncers). Excluding period bouncers, we find perfect agreement between theory and observations. This implies that either we still need to find large numbers of period bouncers within 150\,pc or the observed CVs are much older than our models suggest, i.e. the time-scale for CV formation might be longer and therefore CVs had not enough time yet to evolve past the period minimum. This might again be related to our ignorance of the strength and dependencies of angular momentum loss through magnetic braking.

\section*{Acknowledgements}

We would like to thank an anonymous 
referee for the comments and 
suggestions that helped to improve 
this manuscript.
This work has made use of data from the European Space Agency (ESA) mission {\it Gaia} (\url{https://www.cosmos.esa.int/gaia}), processed by the {\it Gaia} Data Processing and Analysis Consortium (DPAC, {\url{https://www.cosmos.esa.int/web/gaia/dpac/consortium}}). Funding for the DPAC has been provided by national institutions, in particular the institutions participating in the {\it Gaia} Multilateral Agreement.
We thank MCTIC/FINEP (CT-INFRA grant 0112052700) and the Embrace Space Weather Program for the computing facilities at the National Institute for Space Research, Brazil.
DB was supported by the grants \#2017/14289-3 and \#2018/23562-8, S\~ao Paulo Research Foundation (FAPESP).
MRS acknowledges financial support from FONDECYT grant number 1181404. 
BTG was supported by the UK STFC grant ST/P000495.
MZ acknowledges support from CONICYT PAI (Concurso Nacional de Inserci\'on en la Academia 2017, Folio 79170121) and CONICYT/FONDECYT (Programa de Iniciaci\'on, Folio 11170559).
CVR acknowledges CNPq (Proc. 303444/2018-5) and Proc. 2013/26258-4, Funda\c c\~ao de Amparo \`a Pesquisa do Estado de S\~ao Paulo (FAPESP).

%%%%%%%%%%%%%%%%%%%%%%%%%%%%%%%%%%%%
%%%%%% REFERENCES %%%%%%%%%%%%%%%%%%
%%%%%%%%%%%%%%%%%%%%%%%%%%%%%%%%%%%%

\bibliographystyle{mnras}
\bibliography{references} 

%%%%%%%%%%%%%%%%%%%%%%
%%%% APPENDICES %%%%%%
\appendix

\section{Comparison with \ww}
\label{compWW02}

\begin{figure}
   \begin{center}
    \includegraphics[width=0.99\linewidth]
    {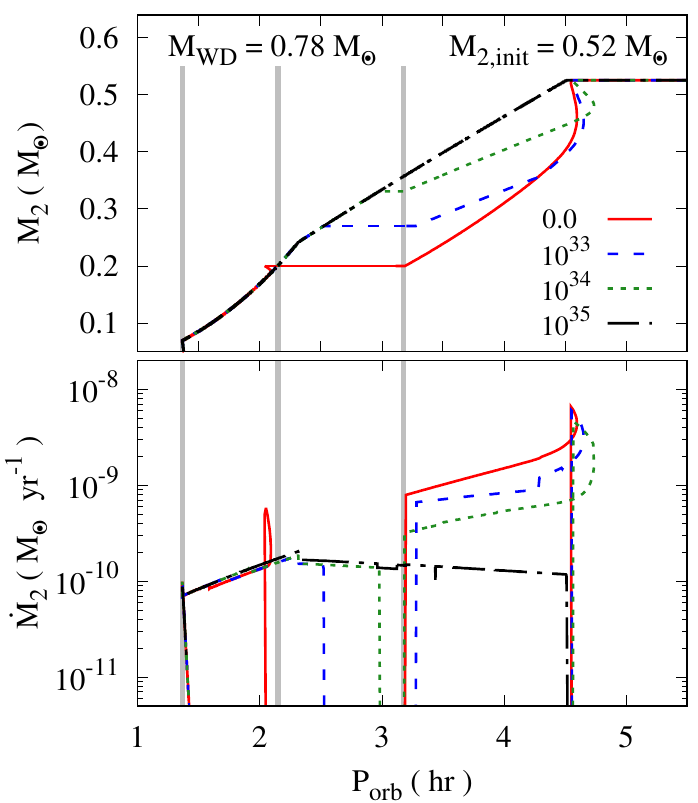}
    \end{center}
  \caption{Evolution with orbital 
  period ($P_{\rm orb}$) of donor mass 
  ($M_2$, top panel), and 
  mass transfer rate 
  ($\dot{M}_2$, bottom panel).
  We show the evolution of one 
  illustrative CV with initial donor 
  and WD masses of 0.52 and 0.78 M$_\odot$,
  respectively, considering four values
  of $\mu_{\rm WD}$, in units of G cm$^3$:
  0.0 (red solid line), 
  $10^{33}$ (long dashed blue line), 
  $10^{34}$ (short dashed green line) 
  and $10^{35}$ (dash-dotted black line).
  We also show in the plots the 
  observational location of the 
  period minimum
  \citep{Gansicke_2009} and 
  gap edges
  \citep{Knigge_2006}, 
  as vertical gray lines. 
  Note that our CV evolution is in good 
  agreement with \ww.
  }
  \label{Fig03}
\end{figure}

In order to validate our approach
for polar evolution,
we compare here
the evolution of particular CVs 
with fig.~8 in \ww, 
by assuming four different values 
of the WD magnetic moment 
$\mu_{\rm WD}=$
0.0, 
$10^{33}$, 
$10^{34}$ 
and 
$10^{35}$~G\,cm$^3$. 
In all 
%four 
cases, the WD mass is 
$0.78$~\Msun~and the initial donor 
mass is $0.52$~\Msun.
Fig. \ref{Fig03} shows the 
evolution of these systems in 
the donor mass vs. orbital period plane 
(top panel) 
and 
mass transfer rate vs. orbital period plane
(bottom panel). 

Starting with the standard non-magnetic 
case (i.e. $\mu_{\rm WD}=0$), 
we first note that the orbital period 
decreases due to MB during the
detached phase when the systems are still progenitors of CVs
(horizontal line) and 
the onset of mass transfer occurs 
at $\approx4.5$~h
(i.e. CV phase starts).
At this point, the donor is driven out 
of thermal equilibrium, 
which leads to a significant 
inflation and an increase in the 
orbital period.
After reaching the CV donor mass-radius 
relation above the period gap, 
the orbital period starts to decrease 
until MB braking is disrupted, 
when the donor becomes fully convective.
This occurs when the donor
mass is $\approx0.2$~\Msun~at a period 
of $\approx$~3~h. 
At this point, the donor has time to 
re-establish thermal equilibrium and
therefore its radius decreases below the Roche-radius:  
the binary becomes detached.
As GR still operates and removes angular 
momentum, the orbital period 
keeps decreasing. This phase of detached 
evolution corresponds to the 
orbital period gap (vertical line 
between $\approx2$ and $\approx3$~h).
When the orbital period is $\approx2$~h,
the donor starts filling its 
Roche lobe again and mass transfer restarts 
(i.e. binary becomes a CV again).
The orbital period further decreases 
until the thermal time-scale exceeds
the mass loss time-scale.
This happens when 
$M_2\approx0.07$~\Msun~and 
$P_{\rm orb}\approx1.36$~h.
After that, the CV is a period bouncer, 
with a degenerate donor,
which causes the period to slightly 
increase in response to further mass loss 
and AML.

As  described  in  Section  \ref{mag}, 
in  polars, the WD magnetic field can 
close field lines from the donor, 
reducing in turn the wind zone,
and consecutively diminishing wind-driven 
AML (i.e. MB).
The effect of increasing $\mu_{\rm WD}$ 
on CV evolution is also shown in 
Fig. \ref{Fig03}.
As the MB efficiency becomes smaller 
with increasing $\mu_{\rm WD}$, 
mass transfer rates above the period 
gap also becomes smaller, 
and the donor star is driven less out of  
thermal equilibrium.
As the donor is therefore less bloated, 
its contraction when MB is disrupted is 
less significant and the systems needs 
less time to become semidetached again. 
In other words, the lower edge of the 
gap is located at longer periods as the 
donor is more massive while entering the 
gap.
As a consequence, the orbital period 
gap becomes partially/completely filled
as $\mu_{\rm WD}$ increases.
For sufficiently high values of 
$\mu_{\rm WD}$, MB is fully
suppressed, and the system evolves 
as an accreting system through  the 
period  gap. 

Comparing our results with those 
from \ww, we find a good agreement.
Indeed, in both cases, the orbital period
gap phase starts at roughly the same
period for all values of $\mu_{\rm WD}$
($\sim3$~h),
the gap width is consistently
reduced for larger $\mu_{\rm WD}$  
and the lower edge of the gap is located
at longer orbital periods.
In fact, as $\mu_{\rm WD}$ becomes 
greater, the MB efficiency becomes
smaller, the mass transfer rate
decreases, and consequentially the 
donor is less bloated.
Given that our evolutionary tracks 
of polars resemble those of \ww, 
we can conclude that we successfully 
incorporated polar evolution in {\sc bse}.

\section{Orbital Period and Distances
for the SDSS systems}

Table~\ref{TabOBS} presents our
SDSS CV sample. It is composed of
systems with reliably determined 
orbital periods from the literature.

%\includepdf[pages=-]{TableSDSS.pdf}

\begin{table*}
\centering
\caption{CVs from SDSS 
with reliable orbital period measurements.
Distances are 
from 
\citet{Bailer_2018}
and have been
derived from Gaia Data Release 2
parallax measurements
\citep{Gaia_2016,Gaia_2018}.
CVs are separated into four types, 
namely 
`NON-MAG' (non-magnetic CVs),
`POLAR' (polars),
`IP' (intermediate polars) and
`CV' (unknown CV type).
}
\label{TabOBS}
\begin{adjustbox}{max width=\linewidth}
\noindent
\begin{threeparttable}
\noindent
\begin{tabular}{l|cccc|cc|l|cccc}
\hline
\hline
SDSSJ &
$P_{\rm orb}$ (min)  & 
Distance (pc) &
Type & 
Reference\tnote{a} & & &
SDSSJ &
$P_{\rm orb}$ (min)  & 
Distance (pc) &
Type & 
Reference\tnote{a}\\
\hline
\hline 
001153.08$-$064739.2  &  144.4  & $ 506.8^{+38.4  }_{-33.5  }$ &   NON-MAG   & $( 1 )$ & & &
091945.10$+$085710.0  &  81.3  & $ 190.7^{+9.5  }_{-8.6  }$ &   NON-MAG   & $( 36 )$ \\
001538.25$+$263656.7  &  146.2  & $ 602.6^{+79.0  }_{-62.9  }$ &   NON-MAG   & $( 2 )$  & & &
092009.54$+$004244.8  &  212.9  & $ 1559.6^{+627.9  }_{-397.7  }$ &   NON-MAG   & $( 13 )$  \\
001856.93$+$345444.2  &  855.0  & $ 2050.9^{+418.7  }_{-307.6  }$ &   CV   & $( 108 )$  & & &
092122.84$+$203857.1  &  84.2   & \tnote{b} &  POLAR & $( 13 )$ \\ 
002728.01$-$010828.5  &  85.4   & \tnote{b} &  NON-MAG & $( 8 )$ & & & 
092229.26$+$330743.6  &  89.1  & $ 569.3^{+221.0  }_{-128.6  }$ &   NON-MAG   & $( 117 )$  \\
003640.29$+$230831.3  &  95.8  & $ 634.7^{+120.9  }_{-88.5  }$ &   NON-MAG   & $( 2 )$  & & &
092444.48$+$080150.9  &  131.2  & $ 616.3^{+285.0  }_{-154.9  }$ &   CV   & $( 15 )$  \\ 
003827.04$+$250925.0  &  136.1  & $ 478.5^{+97.8  }_{-70.0  }$ &   NON-MAG   & $( 3 )$ & & &
093214.82$+$495054.7  &  602.5  & $ 2236.1^{+320.6  }_{-254.8  }$ &   NON-MAG   & $( 13 )$  \\
003941.06$+$005427.5  &  91.4  & $ 977.8^{+693.6  }_{-425.6  }$ &   NON-MAG   & $( 4 )$  & & &
093249.57$+$472523.0  &  95.3  & $ 795.3^{+288.9  }_{-175.6  }$ &   NON-MAG   & $( 37 )$  \\
004335.13$-$003729.8  &  82.3  & $ 345.4^{+101.6  }_{-64.6  }$ &   NON-MAG   & $( 5 )$ & & &
093537.46$+$161950.8  &  92.2  & $ 1292.9^{+567.4  }_{-351.6  }$ &   CV   & $( 43 )$ \\
005050.87$+$000912.7  &  80.3   & \tnote{b} &  NON-MAG & $( 6 )$  & & &
093836.98$+$071455.1  &  269.0  & $ 499.7^{+51.2  }_{-42.7  }$ &   NON-MAG   & $( 38 )$  \\
013132.38$-$090122.2  &  81.5  & $ 301.2^{+25.6  }_{-21.9  }$ &   NON-MAG   & $( 6 )$ & & & 
094325.90$+$520128.8  &  94.6  & $ 798.6^{+221.5  }_{-146.6  }$ &   NON-MAG   & $( 114 )$  \\
013701.06$-$091234.8  &  79.7  & $ 279.1^{+19.2  }_{-16.9  }$ &   NON-MAG   & $( 7 )$  & & &
094431.70$+$035805.5  &  214.8  & \tnote{b} &  NON-MAG & $( 109 )$ \\
015151.86$+$140047.1  &  118.7  & $ 665.2^{+443.3  }_{-215.4  }$ &   NON-MAG   & $( 8 )$ & & & 
094558.24$+$292252.2  &  92.0  & $ 459.9^{+133.9  }_{-86.0  }$ &   POLAR   & $( 43 )$ \\
015543.39$+$002807.1  &  87.1  & $ 317.2^{+14.6  }_{-13.4  }$ &   POLAR   & $( 96 )$  & & &
094636.59$+$444644.7  &  123.6  & $ 386.5^{+65.0  }_{-49.0  }$ &   NON-MAG   & $( 39 )$  \\
023322.60$+$005059.5  &  96.1  & $ 614.4^{+438.7  }_{-206.5  }$ &   CV   & $( 34 )$ & & & 
100515.38$+$191107.9  &  107.6  & $ 388.3^{+41.9  }_{-34.6  }$ &   NON-MAG   & $( 13 )$ \\
032855.00$+$052254.1  &  122.1  & $ 579.6^{+324.8  }_{-163.5  }$ &   POLAR   & $( 77 )$ & & & 
100658.40$+$233724.4  &  267.7  & $ 772.7^{+217.4  }_{-143.1  }$ &   NON-MAG   & $( 40 )$  \\
033449.86$-$071047.8  &  104.0  & $ 428.9^{+66.8  }_{-51.3  }$ &   NON-MAG   & $( 9 )$  & & &
101323.64$+$455858.9  &  118.1  & $ 695.6^{+263.1  }_{-156.3  }$ &   NON-MAG   & $( 36 )$ \\
040714.78$-$064425.1  &  245.0  & $ 659.0^{+41.7  }_{-37.1  }$ &   NON-MAG   & $( 10 )$ & & & 
101534.65$+$090441.9  &  79.9  & $ 262.4^{+12.7  }_{-11.6  }$ &   POLAR   & $( 79 )$ \\
072910.68$+$365838.2  &  150.0  & $ 844.7^{+709.3  }_{-320.9  }$ &   POLAR   & $( 16 )$  & & &
101947.26$+$335753.6  &  92.7  & $ 715.7^{+169.8  }_{-117.4  }$ &   NON-MAG   & $( 41 )$  \\
073208.11$+$413008.7  &  110.9  & \tnote{b} &  NON-MAG & $( 76 )$  & & &
102026.48$+$530433.1  &  97.9  & $ 365.3^{+15.7  }_{-14.5  }$ &   NON-MAG   & $( 42 )$ \\
073817.74$+$285519.6  &  127.0  & $ 767.4^{+308.9  }_{-177.2  }$ &   NON-MAG   & $( 11 )$ & & &
102320.27$+$440509.8  &  97.8  & $ 631.2^{+159.0  }_{-107.7  }$ &   NON-MAG   & $( 43 )$  \\
074531.91$+$453829.3  &  79.5  & $ 375.9^{+69.9  }_{-51.2  }$ &   NON-MAG   & $( 12 )$  & & &
102637.04$+$475426.4  &  96.0  & $ 566.9^{+357.4  }_{-171.2  }$ &   NON-MAG   & $( 44 )$ \\
074640.62$+$173412.8  &  93.5  & $ 952.0^{+867.4  }_{-416.0  }$ &   NON-MAG   & $( 13 )$ & & &
102800.08$+$214813.5  &  210.3  & $ 1047.5^{+190.6  }_{-142.1  }$ &   NON-MAG   & $( 36 )$  \\
074813.54$+$290509.0  &  150.0  & $ 1732.0^{+861.1  }_{-554.8  }$ &   NON-MAG   & $( 14 )$  & & &
102905.21$+$485515.2  &  91.3  & $ 143.3^{+5.1  }_{-4.8  }$ &   NON-MAG   & $( 2 )$ \\
075059.97$+$141150.1  &  134.2  & $ 760.8^{+231.5  }_{-146.9  }$ &   NON-MAG   & $( 15 )$ & & &
103147.99$+$085224.3  &  131.3  & $ 1130.0^{+324.0  }_{-215.2  }$ &   NON-MAG   & $( 115 )$  \\
075117.09$+$144423.5  &  311.6  & $ 841.9^{+181.0  }_{-128.3  }$ &   IP   & $( 98 )$  & & &
103533.03$+$055158.4  &  82.1  & $ 209.2^{+15.6  }_{-13.6  }$ &   NON-MAG   & $( 45 )$ \\
075240.44$+$362823.2  &  164.4  & $ 497.5^{+626.0  }_{-207.1  }$ &   POLAR   & $( 16 )$ & & & 
104051.24$+$151133.7  &  337.7  & $ 1358.7^{+212.7  }_{-164.6  }$ &   CV   & $( 102 )$  \\
075443.00$+$500729.2  &  206.0  & $ 1448.0^{+420.1  }_{-280.7  }$ &   NON-MAG   & $( 6 )$  & & &
104356.65$+$580731.5  &  106.4  & $ 177.7^{+3.3  }_{-3.2  }$ &   NON-MAG   & $( 46 )$  \\
075507.69$+$143547.4  &  84.8  & $ 259.0^{+18.2  }_{-16.0  }$ &   NON-MAG   & $( 13 )$ & & & 
105135.09$+$540435.6  &  114.5  & $ 710.1^{+178.2  }_{-121.1  }$ &   POLAR   & $( 80 )$ \\
075653.11$+$085831.8  &  197.3  & $ 1150.8^{+138.0  }_{-112.0  }$ &   NON-MAG   & $( 17 )$  & & &
105430.43$+$300610.1  &  96.7  & $ 322.0^{+12.6  }_{-11.7  }$ &   NON-MAG   & $( 47 )$ \\ 
075853.03$+$161645.1  &  86.1  & $ 207.6^{+2.5  }_{-2.5  }$ &   IP   & $( 99 )$ & & & 
105550.08$+$095620.4  &  233.9  & $ 877.7^{+147.7  }_{-111.9  }$ &   NON-MAG   & $( 2 )$  \\
075939.78$+$191417.2  &  188.4  & $ 1228.8^{+367.0  }_{-239.2  }$ &   CV   & $( 13 )$  & & &
105656.96$+$494118.3  &  100.2  & $ 311.8^{+13.8  }_{-12.7  }$ &   NON-MAG   & $( 48 )$ \\
080215.39$+$401047.1  &  221.6  & $ 1228.8^{+127.5  }_{-106.3  }$ &   NON-MAG   & $( 13 )$ & & & 
105754.25$+$275947.5  &  90.4  & $ 620.1^{+372.5  }_{-183.3  }$ &   NON-MAG   & $( 43 )$  \\
080303.90$+$251627.0  &  102.0  & $ 938.7^{+512.0  }_{-271.2  }$ &   NON-MAG   & $( 18 )$  & & &
110014.72$+$131552.1  &  94.5  & $ 460.3^{+77.0  }_{-58.1  }$ &   NON-MAG   & $( 13 )$ \\
080434.13$+$510349.2  &  85.0  & $ 145.2^{+3.3  }_{-3.2  }$ &   NON-MAG   & $( 19 )$ & & & 
110425.64$+$450313.9  &  114.8  & $ 320.3^{+14.8  }_{-13.6  }$ &   POLAR   & $( 81 )$ \\
080534.49$+$072029.1  &  329.3  & $ 1663.5^{+570.8  }_{-363.8  }$ &   NON-MAG   & $( 49 )$  & & &
110539.76$+$250628.6  &  113.9  & $ 112.8^{+1.1  }_{-1.1  }$ &   POLAR   & $( 82 )$ \\
080846.19$+$313106.0  &  296.4  & $ 1629.5^{+666.1  }_{-424.6  }$ &   NON-MAG   & $( 13 )$ & & & 
111126.82$+$571238.9  &  55.4  & $ 568.9^{+46.2  }_{-39.9  }$ &   NON-MAG   & $( 111 )$  \\
080908.39$+$381406.2  &  193.0  & $ 1222.2^{+77.6  }_{-69.1  }$ &   NON-MAG   & $( 20 )$  & & &
111544.50$+$425822.4  &  115.9  & $ 101.0^{+1.3  }_{-1.2  }$ &   POLAR   & $( 50 )$  \\
081207.63$+$131824.4  &  116.8  & $ 1150.4^{+793.7  }_{-442.1  }$ &   NON-MAG   & $( 13 )$  & & & 
111721.92$+$520501.0  &  1636.7  & $ 1735.5^{+655.2  }_{-440.7  }$ &   CV   & $( 101 )$ \\
081256.85$+$191157.8  &  230.6  & $ 751.8^{+41.9  }_{-37.8  }$ &   NON-MAG   & $( 21 )$  & & &
112003.39$+$663632.4  &  98.4   & \tnote{b} &  NON-MAG & $( 116 )$   \\
081321.91$+$452809.3  &  416.2  & $ 1544.4^{+383.5  }_{-267.9  }$ &   NON-MAG   & $( 22 )$  & & & 
112253.32$-$111037.5  &  65.2   & $ 811.2^{+731.5  }_{-337.0  }$ &  NON-MAG & $( 112 )$   \\
081352.02$+$281317.2  &  175.1  & $ 1190.0^{+352.0  }_{-231.4  }$ &   NON-MAG   & $( 13 )$  & & &
113122.39$+$432238.5  &  91.1  & $ 338.7^{+21.0  }_{-18.7  }$ &   NON-MAG   & $( 36 )$  \\
081610.83$+$453010.1  &  301.8  & $ 974.8^{+605.4  }_{-326.4  }$ &   NON-MAG   & $( 23 )$  & & & 
113215.50$+$624900.4  &  99.2  & $ 947.9^{+208.7  }_{-148.0  }$ &   NON-MAG   & $( 36 )$   \\
082051.06$+$493432.1  &  99.4  & $ 891.0^{+603.0  }_{-314.0  }$ &   POLAR   & $( 78 )$  & & &
113722.20$+$014858.7  &  109.6  & $ 280.8^{+21.9  }_{-19.0  }$ &   NON-MAG   & $( 42 )$  \\
082236.05$+$510524.5  &  224.5  & $ 765.9^{+31.5  }_{-29.2  }$ &   NON-MAG   & $( 24 )$  & & &
113826.82$+$032207.0  &  84.7  & $ 127.5^{+1.1  }_{-1.1  }$ &   NON-MAG   & $( 51 )$   \\
082409.72$+$493124.4  &  95.0  & $ 1056.4^{+697.6  }_{-403.1  }$ &   NON-MAG   & $( 8 )$  & & & 
113950.57$+$455817.7  &  121.4  & $ 540.3^{+231.5  }_{-129.0  }$ &   NON-MAG   & $( 44 )$  \\
083619.14$+$212105.3  &  105.6  & $ 525.0^{+84.9  }_{-64.6  }$ &   NON-MAG   & $( 25 )$  & & &
114955.68$+$284507.2  &  90.1  & $ 331.5^{+38.8  }_{-31.5  }$ &   POLAR   & $( 83 )$ \\
083642.74$+$532838.0  &  81.8  & $ 161.9^{+2.2  }_{-2.1  }$ &   NON-MAG   & $( 26 )$  & & & 
115207.00$+$404947.6  &  97.5  & $ 739.1^{+521.8  }_{-254.4  }$ &   NON-MAG   & $( 31 )$   \\
083845.23$+$491055.5  &  99.7  & $ 751.1^{+164.7  }_{-116.4  }$ &   NON-MAG   & $( 13 )$  & & &
115215.80$+$491441.7  &  90.2  & $ 318.5^{+19.8  }_{-17.7  }$ &   NON-MAG   & $( 42 )$  \\
083931.35$+$282824.0  &  109.0  & $ 600.9^{+607.0  }_{-252.5  }$ &   NON-MAG   & $( 27 )$ & & &
121209.31$+$013627.7  &  88.4  & $ 149.2^{+4.1  }_{-3.9  }$ &   POLAR   & $( 84 )$ \\ %===> PRE OR AM
084303.98$+$275149.6  &  84.6  & $ 183.7^{+11.2  }_{-10.0  }$ &   NON-MAG   & $( 28 )$   & & &
121607.03$+$052013.9  &  98.8  & $ 448.1^{+531.3  }_{-169.9  }$ &   NON-MAG   & $( 34 )$  \\
084400.10$+$023919.3  &  298.1  & $ 894.4^{+230.7  }_{-155.7  }$ &   NON-MAG   & $( 13 )$ & & &
121913.04$+$204938.3  &  85.5  & $ 281.8^{+36.3  }_{-28.9  }$ &   NON-MAG   & $( 113 )$ \\
084617.12$+$245344.1  &  263.2  & $ 1315.6^{+581.9  }_{-356.2  }$ &   CV   & $( 101 )$ & & & 
122740.83$+$513924.9  &  90.6  & $ 362.4^{+29.2  }_{-25.2  }$ &   NON-MAG   & $( 52 )$  \\
085107.38$+$030834.3  &  93.9  & $ 707.8^{+172.5  }_{-117.8  }$ &   NON-MAG   & $( 13 )$   & & &
123813.73$-$033932.9  &  80.5  & $ 168.6^{+4.6  }_{-4.3  }$ &   NON-MAG   & $( 53 )$ \\
085344.17$+$574840.6  &  97.8  & $ 151.9^{+1.5  }_{-1.5  }$ &   NON-MAG   & $( 29 )$ & & &
123931.98$+$210806.2  &  125.3  & $ 219.4^{+3.9  }_{-3.8  }$ &   NON-MAG   & $( 33 )$ \\
085414.02$+$390537.2  &  113.3  & $ 553.0^{+163.9  }_{-104.9  }$ &   POLAR   & $( 8 )$ & & & 
124417.87$+$300400.8  &  111.5  & $ 643.8^{+363.2  }_{-180.0  }$ &   NON-MAG   & $( 36 )$ \\
085521.17$+$111815.3  &  93.7  & $ 586.8^{+75.2  }_{-60.2  }$ &   NON-MAG   & $( 30 )$   & & &
124426.25$+$613514.6  &  142.9  & $ 680.5^{+87.7  }_{-70.2  }$ &   NON-MAG   & $( 8 )$ \\
085909.18$+$053654.5  &  143.8  & $ 437.1^{+32.4  }_{-28.3  }$ &   POLAR   & $( 13 )$  & & &
125023.79$+$665525.5  &  84.6  & $ 463.2^{+36.4  }_{-31.5  }$ &   NON-MAG   & $( 8 )$ \\
090016.55$+$430118.1  &  301.5  & $ 817.2^{+116.2  }_{-91.3  }$ &   NON-MAG   & $( 13 )$  & & &
125044.42$+$154957.3  &  86.3  & $ 128.0^{+3.0  }_{-2.9  }$ &   POLAR   & $( 85 )$ \\  %===> PRE OR AM
090103.93$+$480911.1  &  112.1  & $ 563.3^{+299.1  }_{-152.8  }$ &   NON-MAG   & $( 8 )$  & & &
125637.10$+$263643.2  &  94.8  & $ 377.6^{+38.5  }_{-32.1  }$ &   NON-MAG   & $( 54 )$ \\
090350.72$+$330036.1  &  85.1  & $ 400.3^{+77.8  }_{-56.4  }$ &   NON-MAG   & $( 31 )$  & & &
130753.86$+$535130.5  &  79.7  & $ 649.4^{+53.6  }_{-46.1  }$ &   POLAR   & $( 86 )$ \\
090403.48$+$035501.2  &  86.0  & $ 277.7^{+77.0  }_{-49.8  }$ &   NON-MAG   & $( 32 )$   & & &
131223.48$+$173659.1  &  91.9  & $ 574.6^{+354.6  }_{-165.1  }$ &   POLAR   & $( 87 )$  \\
090628.25$+$052656.9  &  215.6  & $ 937.5^{+251.5  }_{-168.3  }$ &   NON-MAG   & $( 113 )$ & & &
132411.57$+$032050.5  &  158.7  & \tnote{b} &  POLAR & $( 43 )$ \\
090950.53$+$184947.4  &  252.6  & $ 262.0^{+4.8  }_{-4.6  }$ &   NON-MAG   & $( 33 )$   & & &
132723.38$+$652854.2  &  196.8  & $ 1570.3^{+468.6  }_{-314.3  }$ &   NON-MAG   & $( 55 )$ \\
091127.36$+$084140.7  &  295.7  & $ 964.5^{+547.1  }_{-296.8  }$ &   NON-MAG   & $( 34 )$ & & &
133309.19$+$143706.9  &  132.0  & $ 1634.9^{+994.7  }_{-630.9  }$ &   POLAR   & $( 88 )$ \\ 
091216.20$+$505353.8  &  78.6  & $ 673.6^{+38.2  }_{-34.4  }$ &   NON-MAG   & $( 35 )$   & & &
133940.98$+$484727.9  &  82.5  & $ 149.5^{+2.0  }_{-1.9  }$ &   NON-MAG   & $( 56 )$ \\
091242.18$+$620940.1  &  115.4  & $ 300.4^{+42.0  }_{-33.0  }$ &   NON-MAG   & $( 36 )$  & & &
134323.16$+$150916.8  &  92.7  & $ 393.7^{+27.7  }_{-24.3  }$ &   NON-MAG   & $( 41 )$ \\
091650.76$+$284943.1  &  265.7  & $ 983.3^{+595.2  }_{-336.7  }$ &   NON-MAG   & $( 36 )$   & & &
143317.78$+$101123.3  &  78.1  & $ 223.8^{+10.7  }_{-9.8  }$ &   NON-MAG   & $( 52 )$ \\
\hline
\hline 
\end{tabular}
\end{threeparttable}
\end{adjustbox}
\end{table*}

%%%%%%%%%%%%%%%%%%%%%%%%%%%%%%%%%
%%%%%%%%%%%%%%%%%%%%%%%%%%%%%%%%%
%%%%%%%%%%%%%%%%%%%%%%%%%%%%%%%%%
%%%%%%%%%%%%%%%%%%%%%%%%%%%%%%%%%
%%%%%%%%%%%%%%%%%%%%%%%%%%%%%%%%%
%%%%%%%%%%%%%%%%%%%%%%%%%%%%%%%%%
%%%%%%%%%%%%%%%%%%%%%%%%%%%%%%%%%
%%%%%%%%%%%%%%%%%%%%%%%%%%%%%%%%%
%%%%%%%%%%%%%%%%%%%%%%%%%%%%%%%%%
%%%%%%%%%%%%%%%%%%%%%%%%%%%%%%%%%

\begin{table*}
\centering
\textbf{Table \ref{TabOBS}} -- continued.\\
\begin{adjustbox}{max width=\linewidth}
\noindent
\begin{threeparttable}
\noindent
\begin{tabular}{l|cccc|cc|l|cccc}
\hline
\hline
SDSSJ &
$P_{\rm orb}$ (min)  & 
Distance (pc) &
Type & 
Reference\tnote{a} & & &
SDSSJ &
$P_{\rm orb}$ (min)  & 
Distance (pc) &
Type & 
Reference\tnote{a}\\
\hline
\hline
143500.22$-$004606.3  &  104.7  & $ 558.7^{+205.5  }_{-119.3  }$ &   NON-MAG   & $( 57 )$  & & & 
163722.21$-$001957.1  &  97.0  & $ 2504.0^{+2592.3  }_{-1367.4  }$ &   NON-MAG   & $( 5 )$ \\
145758.21$+$514807.9  &  77.9  & $ 585.2^{+146.8  }_{-99.0  }$ &   NON-MAG   & $( 58 )$ & & & 
164248.52$+$134751.4  &  113.6  & $ 533.8^{+36.0  }_{-31.8  }$ &   NON-MAG   & $( 5 )$ \\
150137.22$+$550123.4  &  81.9  & $ 342.1^{+37.3  }_{-30.7  }$ &   NON-MAG   & $( 31 )$  & & & 
165359.05$+$201010.4  &  90.9  & $ 899.9^{+173.1  }_{-125.8  }$ &   NON-MAG   & $( 36 )$ \\ 
150240.97$+$333423.8  &  84.8  & $ 185.8^{+3.1  }_{-3.0  }$ &   NON-MAG   & $( 31 )$ & & & 
165658.12$+$212139.3  &  97.0  & $ 523.7^{+47.3  }_{-40.2  }$ &   NON-MAG   & $( 36 )$ \\
150441.76$+$084752.6  &  116.5  & $ 1061.6^{+642.4  }_{-306.8  }$ &   NON-MAG   & $( 36 )$  & & & 
165837.70$+$184727.4  &  98.0  & $ 415.1^{+82.6  }_{-59.3  }$ &   NON-MAG   & $( 5 )$ \\
150722.15$+$523040.2  &  66.6  & $ 210.1^{+6.5  }_{-6.1  }$ &   NON-MAG   & $( 52 )$ & & & 
165951.69$+$192745.6  &  203.0  & $ 1573.3^{+224.1  }_{-175.6  }$ &   NON-MAG   & $( 36 )$ \\
151302.29$+$231508.4  &  140.4  & $ 756.7^{+30.8  }_{-28.5  }$ &   NON-MAG   & $( 42 )$  & & & 
170053.29$+$400357.6  &  116.4  & $ 524.1^{+34.2  }_{-30.3  }$ &   POLAR   & $( 16 )$ \\
151415.65$+$074446.4  &  88.7  & $ 181.1^{+7.9  }_{-7.2  }$ &   POLAR   & $( 85 )$ & & & %===> PRE OR AM
170213.25$+$322954.1  &  114.1  & $ 292.3^{+11.3  }_{-10.5  }$ &   NON-MAG   & $( 70 )$ \\
152419.33$+$220920.0  &  93.6  & $ 469.2^{+79.5  }_{-59.6  }$ &   NON-MAG   & $( 15 )$ & & & 
171145.08$+$301319.9  &  80.3  & $ 364.1^{+101.9  }_{-65.5  }$ &   NON-MAG   & $( 8 )$ \\
152613.96$+$081802.3  &  107.4  & $ 378.3^{+43.6  }_{-35.5  }$ &   NON-MAG   & $( 59 )$  & & &  
173008.38$+$624754.7  &  110.3  & $ 528.5^{+10.8  }_{-10.4  }$ &   NON-MAG   & $( 36 )$ \\
153634.42$+$332851.9  &  132.6  & $ 819.0^{+244.4  }_{-155.8  }$ &   NON-MAG   & $( 36 )$  & & & 
173102.22$+$342633.2  &  300.2  & $ 2112.2^{+893.9  }_{-529.3  }$ &   NON-MAG   & $( 2 )$ \\ 
153817.35$+$512338.0  &  93.1  & $ 622.2^{+73.4  }_{-59.6  }$ &   NON-MAG   & $( 8 )$  & & & 
204448.92$-$045928.8  &  2420.0  & $ 2016.9^{+517.3  }_{-349.9  }$ &   NON-MAG   & $( 71 )$ \\
154104.67$+$360252.8  &  84.3  & $ 414.4^{+16.9  }_{-15.6  }$ &   POLAR   & $( 13 )$  & & & 
204720.76$+$000007.7  &  89.0  & $ 478.9^{+259.3  }_{-124.9  }$ &   NON-MAG   & $( 115 )$ \\
154453.60$+$255348.8  &  361.8  & $ 519.6^{+21.4  }_{-19.8  }$ &   NON-MAG   & $( 60 )$ & & & 
204817.85$-$061044.8  &  87.3  & $ 608.6^{+370.5  }_{-168.5  }$ &   NON-MAG   & $( 72 )$ \\
154539.08$+$142231.4  &  280.8  & $ 3452.5^{+862.7  }_{-604.6  }$ &   NON-MAG   & $( 61 )$  & & & 
204827.91$+$005008.9  &  252.0  & $ 591.8^{+74.5  }_{-59.7  }$ &   POLAR   & $( 93 )$ \\ 
155247.18$+$185629.1  &  113.5  & $ 131.3^{+0.8  }_{-0.8  }$ &   POLAR   & $( 89 )$ & & & 
205017.83$-$053626.8  &  94.2   & \tnote{b} &  POLAR & $( 97 )$ \\
155331.11$+$551614.4  &  263.5  & $ 185.3^{+3.3  }_{-3.2  }$ &   POLAR   & $( 90 )$  & & & 
205914.87$-$061220.4  &  107.5  & $ 2031.9^{+1495.5  }_{-783.0  }$ &   NON-MAG   & $( 6 )$  \\
155412.33$+$272152.4  &  151.9  & $ 207.6^{+3.2  }_{-3.1  }$ &   POLAR   & $( 91 )$ & & & 
210014.11$+$004445.9  &  120.0  & $ 928.0^{+131.9  }_{-103.2  }$ &   NON-MAG   & $( 73 )$ \\
155531.98$-$001054.9  &  113.5  & $ 648.5^{+190.8  }_{-120.7  }$ &   NON-MAG   & $( 6 )$ & & & 
210449.94$+$010545.8  &  103.6  & $ 1999.3^{+1515.9  }_{-927.1  }$ &   NON-MAG   & $( 6 )$ \\
155644.23$-$000950.2  &  115.4  & $ 308.2^{+22.1  }_{-19.4  }$ &   NON-MAG   & $( 36 )$  & & & 
211605.42$+$113407.4  &  80.2   & \tnote{b} &  NON-MAG & $( 8 )$ \\
155654.47$+$210718.9  &  119.8  & $ 319.3^{+10.0  }_{-9.4  }$ &   NON-MAG   & $( 62 )$ & & & 
214140.43$+$050729.9  &  76.0  & $ 257.3^{+39.1  }_{-30.1  }$ &   NON-MAG   & $( 74 )$ \\
155656.92$+$352336.6  &  126.9  & $ 1377.5^{+491.1  }_{-301.9  }$ &   NON-MAG   & $( 63 )$  & & & 
215411.12$-$090121.6  &  319.0  & \tnote{b} &  CV & $( 8 )$  \\
160745.02$+$362320.7  &  225.4  & $ 1815.9^{+764.1  }_{-465.5  }$ &   NON-MAG   & $( 13 )$ & & & 
220553.98$+$115553.7  &  82.8  & $ 871.3^{+878.8  }_{-357.5  }$ &   NON-MAG   & $( 75 )$ \\
161007.50$+$035232.7  &  190.5  & $ 323.3^{+11.8  }_{-11.0  }$ &   POLAR   & $( 36 )$  & & & 
221832.76$+$192520.2  &  129.5  & $ 237.5^{+7.3  }_{-6.9  }$ &   POLAR   & $( 94 )$ \\
161027.61$+$090738.4  &  81.9  & $ 400.8^{+132.9  }_{-80.0  }$ &   NON-MAG   & $( 64 )$  & & & 
223439.93$+$004127.2  &  127.3  & $ 494.8^{+44.5  }_{-37.8  }$ &   NON-MAG   & $( 13 )$ \\
161033.63$-$010223.2  &  80.5  & $ 242.5^{+25.0  }_{-20.8  }$ &   NON-MAG   & $( 65 )$ & & & 
223843.83$+$010820.6  &  194.3  & $ 2172.8^{+866.4  }_{-571.9  }$ &   IP   & $( 5 )$  \\
161909.10$+$135145.5  &  412.6  & $ 1671.8^{+301.7  }_{-224.1  }$ &   NON-MAG   & $( 36 )$  & & & 
224736.37$+$250436.3  &  81.6   & \tnote{b} &  NON-MAG & $( 110 )$ \\
161935.76$+$524631.8  &  100.5  & $ 438.5^{+50.7  }_{-41.3  }$ &   POLAR   & $( 92 )$ & & & 
225831.18$-$094931.7  &  118.9  & $ 294.0^{+5.0  }_{-4.8  }$ &   NON-MAG   & $( 9 )$ \\
162501.74$+$390926.3  &  78.7  & $ 291.5^{+8.6  }_{-8.1  }$ &   NON-MAG   & $( 66 )$ & & & 
230351.64$+$010651.0  &  110.5  & $ 635.8^{+230.4  }_{-136.9  }$ &   NON-MAG   & $( 13 )$ \\
162520.29$+$120308.7  &  138.2  & $ 457.2^{+58.0  }_{-46.4  }$ &   NON-MAG   & $( 67 )$  & & & 
230949.12$+$213516.7  &  255.8  & $ 172.1^{+1.9  }_{-1.8  }$ &   CV   & $( 103 )$ \\
162718.38$+$120434.9  &  150.0  & $ 1152.6^{+512.1  }_{-277.8  }$ &   NON-MAG   & $( 68 )$ & & & 
231930.43$+$261518.6  &  180.6  & $ 516.2^{+27.7  }_{-25.1  }$ &   POLAR   & $( 95 )$ \\
162936.53$+$263519.5  &  134.0  & $ 683.9^{+124.5  }_{-91.8  }$ &   POLAR   & $( 13 )$  & & & 
233325.92$+$152222.1  &  83.1  & $ 745.8^{+239.7  }_{-150.0  }$ &   IP   & $( 100 )$ \\
163545.72$+$112458.0  &  189.0  & $ 1037.8^{+45.9  }_{-42.3  }$ &   NON-MAG   & $( 69 )$  & & & 
  &    &  &   &   \\
\hline
\hline 
\end{tabular}
\begin{tablenotes}
\item[a] References:
(1) \citet{RebassaMansergas_2014},
(2) \citet{Thorstensen_2016},
(3) \citet{Pavlenko_2016},
(4) \citet{Southworth_2010},
(5) \citet{Southworth_2008},
(6) \citet{Southworth_2007},
(7) \citet{Pretorius_2004},
(8) \citet{Dillon_2008},
(9) \citet{Kato_2009},
(10) \citet{Ak_2005},
(11) \citet{Szkody_2003},
(12) \citet{Mukadam_2013},
(13) \citet{Gansicke_2009},
(14) \citet{Szkody_2004},
(15) \citet{Southworth_2010b},
(16) \citet{Homer_2005},
(17) \citet{Tovmassian_2014},
(18) \citet{Szkody_2005},
(19) \citet{Zharikov_2008},
(20) \citet{Gil_2007},
(21) \citet{Gulsecen_2014},
(22) \citet{Thorstensen_2004},
(23) \citet{Shears_2012},
(24) \citet{Stanishev_2006},
(25) \citet{Thorstensen_1997},
(26) \citet{Howell_1988},
(27) \citet{Kato_2014},
(29) \citet{Ringwald_1994},
(30) \citet{Arenas_1998},
(31) \citet{Savoury_2011},
(32) \citet{Woudt_2005},
(33) \citet{Feline_2005},
(34) \citet{Southworth_2006},
(35) \citet{Rutkowski_2009},
(36) \citet{Thorstensen_2015},
(37) \citet{Homer_2006},
(38) \citet{Thorstensen_2001},
(39) \citet{Feline_2004},
(40) \citet{Southworth_2009},
(41) \citet{Aungwerojwit_2006},
(42) \citet{Patterson_2003},
(43) \citet{Southworth_2015},
(44) \citet{Breedt_2014},
(45) \citet{Littlefair_2006},
(46) \citet{Steeghs_2003},
(47) \citet{Wagner_1998},
(48) \citet{Thorstensen_1996},
(49) \citet{Woudt_2012},
(50) \citet{Schmidt_1999},
(51) \citet{Shafter_1984},
(52) \citet{Littlefair_2008},
(53) \citet{Pala_2019a},
(54) \citet{Howell_1995},
(55) \citet{Wolfe_2003},
(56) \citet{Gansicke_2006},
(57) \citet{Feline_2004b},
(58) \citet{Uthas_2012},
(59) \citet{Olech_2003},
(60) \citet{Skinner_2011},
(61) \citet{Ringwald_2005},
(62) \citet{Thorstensen_2002},
(63) \citet{Hardy_2017},
(64) \citet{Kato_2015},
(65) \citet{Woudt_2004},
(66) \citet{Thorstensen_2002b},
(67) \citet{Olech_2011b},
(68) \citet{Shears_2009},
(69) \citet{Gil_2007b},
(70) \citet{Khruzina_2015},
(71) \citet{Peters_2005},
(72) \citet{Woudt_2010},
(73) \citet{Olech_2009},
(74) \citet{Szkody_2014},
(75) \citet{Southworth_2008b},
(76) \citet{Shears_2011},
(77) \citet{Babina_2017},
(78) \citet{Schwope_2002b},
(79) \citet{Burwitz_1998},
(80) \citet{Morris_1987},
(81) \citet{Bidaud_1996},
(82) \citet{Cropper_1986},
(83) \citet{Howell_1995b},
(84) \citet{Burleigh_2006},
(85) \citet{Breedt_2012},
(86) \citet{Katajainen_2000},
(87) \citet{Vogel_2008},
(88) \citet{Schmidt_2008},
(89) \citet{Schwope_1991},
(90) \citet{Szkody_2003b},
(91) \citet{Thorstensen_2002c},
(92) \citet{Denisenko_2016},
(93) \citet{Schmidt_2005},
(94) \citet{Thorstensen_2013},
(95) \citet{Shafter_2008},
(96) \citet{Woudt_2004b},
(97) \citet{Potter_2006},
(98) \citet{Evans_2006},
(99) \citet{Gil_2004},
(100) \citet{Southworth_2007b},
(101) \citet{Drake_2014},
(102) \citet{Abbott_1990},
(103) \citet{Thorstensen_2009},
(104) \citet{Hilton_2009},
(105) \citet{Thorstensen_2005},
(106) \citet{Reimers_1999},
(107) \citet{Schmidt_2007},
(108) \citet{Buitrago_2013},
(109) \citet{Mennickent_2002},
(110) \citet{Kato_2013},
(111) \citet{Kennedy_2015},
(112) \citet{Breedt_2012b},
(113) Schmidtobreick (private communication), % Unpublished VLT spectra by Linda Schmidtobreick,
(114) Thorstensen (private communication), % Unpublished RV study by John Thorstensen,
(115) Breedt (private communication), % Unpublished EFOSC RV study by Elm\'e  Breedt, 
(116) Stein (private communication), % Unpublished superhump period by Bill Stein,
(117) Southworth (private communication). %Unpublished WHT spectra by John Southworth
\item[b] There is no parallax measurement in Gaia DR2. 
\end{tablenotes}
\end{threeparttable}
\end{adjustbox}
\end{table*}

%%%%%%%%%%%%%%%%%%%%%%

% Don't change these lines
\bsp	% typesetting comment
\label{lastpage}

\end{document}